\documentclass[journal=jacsat,manuscript=article]{achemso}

\usepackage[version=3,version=4]{mhchem} % Formula subscripts using \ce{}

\usepackage{epstopdf}% To incorporate .eps illustrations using PDFLaTeX, etc.
\usepackage[caption=false]{subfig}% Support for small, `sub' figures and tables
\usepackage{amsmath,amsfonts,amssymb}
\usepackage{chemmacros}
\DeclareMathOperator{\sgn}{sgn}
\DeclareMathOperator{\coef}{coef}

\usepackage[numbers,sort&compress]{natbib}% Citation support using natbib.sty
\bibpunct[, ]{[}{]}{,}{n}{,}{,}% Citation support using natbib.sty
\usepackage{xcolor}
\usepackage[british]{babel}
\usepackage[normalem]{ulem}

\author{Juan C. Morales}
\affiliation[Icesi]{Department of Pharmaceutical and Chemical Sciences,  Universidad Icesi, Cali, Colombia}
\author{Carlos A. Arango}
\affiliation[Icesi]{Department of Pharmaceutical and Chemical Sciences,  Universidad Icesi, Cali, Colombia}
\email{caarango@icesi.edu.co}

\title[title]
  {Geometric-algebraic approach to aqueous solutions of diprotic acids and its buffer mixtures}


\begin{document}

%%%%%%%%%%%%%%%%%%%%%%%%%%%%%%%%%%%%%%%%%%%%%%%%%%%%%%%%%%%%%%%%%%%%%
%% The "tocentry" environment can be used to create an entry for the
%% graphical table of contents. It is given here as some journals
%% require that it is printed as part of the abstract page. It will
%% be automatically moved as appropriate.
%%%%%%%%%%%%%%%%%%%%%%%%%%%%%%%%%%%%%%%%%%%%%%%%%%%%%%%%%%%%%%%%%%%%%

%%%%%%%%%%%%%%%%%%%%%%%%%%%%%%%%%%%%%%%%%%%%%%%%%%%%%%%%%%%%%%%%%%%%%
%% The abstract environment will automatically gobble the contents
%% if an abstract is not used by the target journal.
%%%%%%%%%%%%%%%%%%%%%%%%%%%%%%%%%%%%%%%%%%%%%%%%%%%%%%%%%%%%%%%%%%%%%
\begin{abstract}
A closed-form analytical expression for $\ce{[H3O+]}$ has been obtained for aqueous solutions of diprotic acids and its soluble salts. This formula allows to calculate the pH of aqueous solutions of diprotic acids, their buffer solutions, and the titrations of these two by a strong base, from the values of p$K_1$, p$K_2$, and the effective concentrations of the acid and the base, $\bar{C}_\mathrm{a}$ and $\bar{C}_\mathrm{b}$ respectively. It is shown that a strong base titration of an acid, or its buffer solutions, is always a linear path in the $\bar{C}_\mathrm{a}$--$\bar{C}_\mathrm{b}$ plane, which allows a simple analysis of the pH stability of buffer solutions. The mathematical analysis of the equilibrium equations of the dissolution of a diprotic acid in water and the physical constraints allowed to obtain two approximate equations for the diprotic acids. One of the approximations is useful for acids with $\mathrm{p}K_2-\mathrm{p}K_1\le\log_{10}4$, the other for acids with $\mathrm{p}K_2-\mathrm{p}K_1\le-\log_{10}4$.

\end{abstract}

%%%%%%%%%%%%%%%%%%%%%%%%%%%%%%%%%%%%%%%%%%%%%%%%%%%%%%%%%%%%%%%%%%%%%
%% Start the main part of the manuscript here.
%%%%%%%%%%%%%%%%%%%%%%%%%%%%%%%%%%%%%%%%%%%%%%%%%%%%%%%%%%%%%%%%%%%%%
\section{Introduction}
Diprotic acids are of central importance in biochemistry, physiology, and industrial and environmental chemistry. In biochemistry, several amino acids behave as diprotic acids with two dissociated protons: one proton on the $\alpha$ amino group and one on the $\alpha$ carboxyl group \cite{Scholz2018}. In physiology, the regulation of blood pH cannot be understood without considering the buffer made by carbonic acid, $\ce{H2CO3}$, and the bicarbonate ion, $\ce{HCO3-}$ \cite{Hamm2015}. In environmental chemistry, the current model for understanding ocean acidification is based on the aqueous chemical equilibrium between $\ce{CO2}$, $\ce{H2CO3}$, $\ce{HCO3-}$, and $\ce { CO3-}$ \cite{Doney2020}.

A Br{\o}nsted diprotic acid is a chemical substance $\ce{H2B}$ that partially dissolves in water producing hydronium ion, $\ce{H3O+}$, and the conjugate base, $\ce{HB-}$. This conjugate base further dissociates partially producing the second conjugate base, $\ce{B^{2-}}$. In the state of equilibrium, the concentrations of the chemical species are constant \cite{Denbigh1981,Burgot2012}. The equilibrium concentrations of the chemical species are given by the equations of chemical equilibrium, and the chemical and electric balance \cite{Skoog2022}. The aqueous dissociation of a diprotic acid and, its soluble salts, involves five chemical species and five mathematical relations between these species, therefore, in principle is possible to obtain the equilibrium concentrations of all the chemical species by solving this system of equations. In practice, the system of equations involves nonlinear terms making difficult to obtain exact mathematical expression for the equilibrium concentrations. For the dissociation of a diprotic acid and its soluble salts, the algebraic manipulation of the system of equations gives a quartic equation for the concentration of $\ce{H3O+}$, $\ce{[H3O+]}$. The equilibrium concentration of the hydronium ion is obtained by finding the roots of its quartic equation. Although there is a quartic formula that gives the explicit roots of a quartic equation, it is not practical to use due to its complexity. The use of the quartic formulas gives four roots, each of them implies the execution of at least 57 mathematical operations. Although these type of calculation is a simple task for modern computers, the formulas obtained from the quartic equation are not simplified which causes accumulation of computational error. On the other hand, graphical and numerical solutions are easily obtained using numerical calculators and computer software \cite{Kalka2021}. Although the graphical-numerical approach is fast and efficient to calculate concentrations as function of the pH, it has some disadvantages against an analytical closed form solution. The analytical solution can be differentiated to study buffers and buffer stability, or can be easily function-composed to study titrations of acids and buffers by strong bases \cite{Caicedo2023}. Another advantage of an analytical closed form is the possibility to analyze mathematically the effect on the pH of parameters such as the concentrations and the acid dissociation constants. In this work it has been found that that the constrain p$K_2-$p$K_1 \ge \log_{10}{4}$, on the p$K$s of the acid, has an important effect on the nature of the roots of the quartic polynomial for $\ce{[H3O+]}$. This constrain has been previously obtained by considering isomerization of the ionization products and a detailed equilibrium scheme in which the micro-equilibrium constants correspond to partial equilibria \cite{Adams1916,CRChandbook2007,Petrossyan2019}. Direct observation of the experimental values of p$K_1$ and p$K_2$ of a large set of diprotic acids allowed to find that several compounds, in particular the nitrogenous organics, follow the constrain p$K_2-$p$K_1 \le -\log_{10}{4}$.

The main result of this paper is a closed form analytical solutions for \ce{[H3O+]} for the full chemical equilibrium of the aqueous dissociation of a diprotic acid, and its monobasic and dibasic salts. The use of effective acid and base concentrations allow to have only one mathematical expression for aqueous dissolutions of diprotic acids, and buffer dissolutions of diprotic acids and its soluble salts. In this work it is shown how this unified approach to diprotic acids and its buffers allows to study the pH stability of buffer solutions in relation with the equivalent acid dissolution. 

This article is organized as follows: In the Theory and Methods section, the first subsection is dedicated to establishing the notation and fundamental equation of chemical equilibrium and physical constraints. In this subsection a unified notation is introduced, allowing the same equations to be used for: aqueous solutions of diprotic acids, buffers of diprotic acids, and titrations with strong bases. In the second subsection of Theory and Methods, a mathematical expression for $\ce{[H3O+]}$ is obtained and analyzed, showing the complete expression for $\ce{[H3O+]}$. The final expressions of $\ce{[H3O+]}$ are written in algebraic terms using basic arithmetic operations and radicals (square and cube roots). These expressions can be used with the help of computer algebra software to obtain the pH without approximations. The reader interested in the mathematical details and the procedures used to obtain these equations is referred to the Appendix. The third and final subsection introduces exact expressions for the titration functions of the aqueous solution and the buffer solution of diprotic acids. The Results and Discussion section shows the results obtained using the expressions obtained previously. The first section shows that although most diprotic acids obey the condition $\mathrm{p}K_2-\mathrm{p}K_1\ge\log_{10}4$, there are some acids that follow the condition $ \mathrm{p}K_2-\mathrm{p}K_1\le -\log_{10}{4}$. In the next subsection, the physical constraints of diprotic acids are used to obtain two approximations for $\ce{[H3O+]}$, these approximations are used to obtain analytical expressions for the upper and lower limits of pH. In the next subsection, we discuss the common approach of monitoring the second dissociation constant and show that in the case of micro-molar concentrations this approach fails. The following subsection shows the use of exact closed forms of pH and titration functions to analyze the neutralization of aqueous solutions of diprotic acids and their buffer mixtures. The differences between the exact expressions of this work and the approximate results of recent works for two cases are shown in detail: maleic acid and 1,8-Octanediamine. Finally, the last subsection of Results and Discussion shows an analysis of the pH stability of diprotic acid buffer solutions. In this subsection, the pH stability is analyzed as a parametric curve in the plane formed by the pH of the acid and the pH of the corresponding buffer solution.

\section{Theory and Methods}\label{sec:theory}

\subsection{Aqueous solutions of weak diprotic acids and their salts}\label{subsec:notation_and_equations}
The aqueous dissociation equilibrium of a diprotic weak acid $\ce{H2B}$ is given by the chemical equations 
\begin{align}
    \ce{H2B + H2O &<--> H3O+ + HB-},\label{ce:HA_diss1}\\
        \ce{HB- + H2O &<--> H3O+ + B^{2-}},\label{ce:HA_diss2}\\
    \ce{2 H2O &<--> H3O+ + OH-}.\label{ce:H2O_auto}
\end{align}
Relevant chemical species are $\ce{H3O+}$, $\ce{OH-}$, $\ce{H2B}$, $\ce{HB-}$, and $\ce{B^{2-}}$ with equilibrium molar concentrations $\ce{[H3O+]}$, $\ce{[OH^-]}$, $\ce{[H2B]}$, $\ce{[HB^-]}$, and $\ce{[B^{2-}]}$, respectively. The equilibria displayed in equations \eqref{ce:HA_diss1}--\eqref{ce:HA_diss2} are effective equilibria since the two protons of $\ce{H2B}$ can dissociate separately and not necessarily consecutively \cite{Adams1916,Scholz2018,Petrossyan2019}. 

A solution of the acid $\ce{H2B}$ is prepared in water at analytical molar concentration $C_\mathrm{a}$. Once the system reaches chemical equilibrium, the concentrations of the chemical species are given by five physical conditions: two weak acid dissociation constant $K_\mathrm{1}$ and $K_\mathrm{2}$, the water auto-ionization constant $K_\mathrm{w}$, the electric neutrality, and the mass balance,
\begin{align}
    K_\mathrm{1}&=\frac{\ce{[H3O+]}}{C^\circ}\frac{\ce{[HB^-]}}{C^\circ}\left(\frac{\ce{[H2B]}}{C^\circ}\right)^{-1},\label{eq:K1a}\\
    K_\mathrm{2}&=\frac{\ce{[H3O+]}}{C^\circ}\frac{\ce{[B^{2-}]}}{C^\circ}\left(\frac{\ce{[HB^-]}}{C^\circ}\right)^{-1},\label{eq:K2a}\\
    K_\mathrm{w}&=\frac{\ce{[H3O^+]}}{C^\circ}\frac{\ce{[OH^-]}}{C^\circ},\label{eq:Kw}\\
    \ce{[H3O+]}&=\ce{[OH^-]}+\ce{[HB^-]}+2\ce{[B^{2-}]},\label{eq:charge_balance}\\
    C_\mathrm{a}&=\ce{[HB]}+\ce{[HB^-]}+\ce{[B^{2-}]},\label{eq:matter_balance}
\end{align}
respectively. The standard molar concentration is ${C^\circ=1\,\mathrm{M}}$. The acid constants $K_\mathrm{1}$ and $K_\mathrm{2}$ are dimensionless, and their value range typically between $10^{-10}$ and $10^{-1}$. In this work, the biochemical standard state $C^\standardstate=C^\circ\sqrt{K_\mathrm{w}}$ is used to define the dimensionless variables:
${x=\ce{[H3O^+]}/C^\standardstate}$, ${y=\ce{[OH^-]}/C^\standardstate}$, ${z_0=\ce{[H2B]}/C^\standardstate}$, ${z_1=\ce{[HB^-]}/C^\standardstate}$, ${z_2=\ce{[B^{2-}]}/C^\standardstate}$, and the parameter  
${c_{\mathrm{a}}=C_\mathrm{a}/C^\standardstate}$. These definitions make the equilibrium constants ${k_1=K_\mathrm{1}/\sqrt{K_\mathrm{w}}}$, ${k_2=K_\mathrm{2}/\sqrt{K_\mathrm{w}}}$, and ${k_\mathrm{w}=1}$. In terms of the new variables and constants, equations \eqref{eq:K1a}--\eqref{eq:matter_balance} are replaced by
\begin{align}
    k_\mathrm{1}&=\frac{xz_1}{z_0},\label{eq:ka1}\\
    k_\mathrm{2}&=\frac{xz_2}{z_1},\label{eq:ka2}\\
    k_\mathrm{w}&=xy=1,\label{eq:kw}\\
    x&=y+z_1+2z_2,\label{eq:cb}\\
    c_\mathrm{a}&=z_0+z_1+z_2.\label{eq:c}
\end{align}
The equations for electric neutrality \eqref{eq:cb} and mass balance \eqref{eq:c} are explicitly affected by the presence of a strong base and salts of the conjugate bases $\ce{HB-}$ and $\ce{B^{2-}}$, \emph{e.g.} $\ce{NaOH}$, $\ce{NaHB}$ and $\ce{Na2B}$ respectively. If the dimensionless concentrations of the strong base and salts are ${c_\mathrm{b}=\ce{[NaOH]}/C^\standardstate}$, ${s_\mathrm{1}=\ce{[NaHB]}/C^\standardstate}$ and ${s_\mathrm{2}=\ce{[Na2B]}/C^\standardstate}$, the charge and mass balance equations are modified to
\begin{align}
    x+\bar{c}_\mathrm{b}&=y+z_1+2z_2,\label{eq:cb_2}\\
    \bar{c}_\mathrm{a}&=z_0+z_1+z_2\label{eq:c_2}, 
\end{align}
with effective concentrations $\bar{c}_\mathrm{a}=c_\mathrm{a}+s_\mathrm{1}+s_\mathrm{2}$ and $\bar{c}_\mathrm{b}=c_\mathrm{b}+s_\mathrm{1}+2s_\mathrm{2}$. These effective dimensionless variables are related to the effective molar concentrations by $\bar{C}_\mathrm{a}=C^\standardstate \bar{c}_\mathrm{a}$, and $\bar{C}_\mathrm{b}=C^\standardstate \bar{c}_\mathrm{b}$.

The use of $y=1/x$, obtained from equation \eqref{eq:kw},  in the equations \eqref{eq:ka1}, \eqref{eq:ka2}, \eqref{eq:cb_2} and \eqref{eq:c_2}, gives a non-linear system of four equation $\mathcal{S}_4$ with four unknowns $x$, $z_0$, $z_1$ and $z_2$. 

\subsection{Mathematical expression for $\ce{[H3O+]}$}\label{subsec:methematical_expressions}

Before obtaining the full solution of the non-linear system $\mathcal{S}_4$ is useful to analyze the linear subsystem $\mathcal{S}_3$ made by equations \eqref{eq:ka1}, \eqref{eq:ka2} and \eqref{eq:c_2}. This subsystem can be easily solved to obtain the concentrations $z_0$, $z_1$, $z_2$, in terms of $x$. The linear system $\mathcal{S}_3$ can be expressed as $\mathsf{K}\bm{z}=\bm{c}$ with $\mathsf{K}=\mathsf{K}(x)$ given by
\begin{equation}\label{eq:matrix_K}
\mathsf{K}=\begin{pmatrix}
    k_1 & -x & 0 \\
    0 & k_2 & -x \\
    1 & 1 & 1
    \end{pmatrix},
\end{equation}
$\bm{z}=\left(z_0,z_1,z_2\right)^\intercal$, and $\bm{c}=\left(0,0,\bar{c}_\mathrm{a}\right)^\intercal$. Solving for $\bm{z}$ gives 
\begin{equation}\label{eq:vector_z}
    \bm{z}=\frac{\bar{c}_\mathrm{a}}{\det{\mathsf{K}}}\begin{pmatrix}
        x^2 \\
        k_1 x \\
        k_1 k_2
    \end{pmatrix},
\end{equation}
with $\det{\mathsf{K}}=x^2 + k_1 x + k_1 k_2$ as the determinant of $\mathsf{K}$. It is convenient to write this determinant as ${\det{\mathsf{K}}=(x-\kappa_1)(x-\kappa_2)}$ with
\begin{equation}
    \kappa_{1,2}=\tfrac{k_1}{2}\left(-1\pm\sqrt{1-4\kappa}\right),
\end{equation}
and $\kappa=k_2/k_1$ as the ratio of the diprotic dissociation constants. These $\kappa_{1,2}$ are related to the Simms constants \cite{Simms1926,Kalka2021}, $g_{1,2}$, by $g_{1,2}=-\kappa_{1,2}$.

Given the condition $\kappa\le 1/4$, \emph{i.e.} $k_1\ge 4k_2$, the roots $\kappa_{1,2}$ are both no positive real numbers, $\kappa_{1,2}\le 0$, otherwise these roots are a pair of complex conjugate numbers with negative real part, \emph{i.e.} $\kappa_1=\kappa_2^*$ and $\mathrm{re}{(\kappa_1)}=\mathrm{re}{(\kappa_2)}<0$. The inequality $k_1\ge 4k_2$ have been obtained previously by Adams in his analysis of polyprotic acid dissociations \cite{Adams1916}. 

The solution of $\mathcal{S}_3$ gives the concentrations $\bm{z}=\left(z_0,z_1,z_2\right)^\intercal$ as functions of $x$, $k_1$, $k_2$ and $\bar{c}_\mathrm{a}$. Although  $\bm{z}$ does not depend explicitly on $\bar{c}_\mathrm{b}$, it depends implicitly through $x$. This dependency is specified by using $y=1/x$ in equation \eqref{eq:cb_2}, which gives   \begin{equation}\label{eq:charge_balance_z}
    x-\tfrac{1}{x}+\bar{c}_\mathrm{b}=-\bm{q}\cdot\bm{z},
\end{equation}
with $\bm{q}=\left(0,-1,-2\right)^\intercal$ as the vector of electric charges of $z_0$, $z_1$, and $z_2$. This equation keeps some similarity with equation (41) used in the work of Kalka \cite{Kalka2021}. However, unlike that article, in this paper closed analytic solutions for $x$ are obtained instead of graphical or numerical solutions.

Multiplying equation \eqref{eq:charge_balance_z}  by $x\det{\mathsf{K}}$, and expanding the scalar product, produces
\begin{equation}\label{eq:factrorized_Pa}
    \left(x-\kappa_1\right)\left(x-\kappa_2\right)\left(x^2+\bar{c}_\mathrm{b}x-1\right)=\bar{c}_\mathrm{a} k_1 x \left(x+2 k_2\right),
\end{equation}
which can be written
\begin{equation}\label{eq:factrorized_Pa_2}
     \left(x-\kappa_1\right)\left(x-\kappa_2\right)\left(x-\sigma_1\right)\left(x-\sigma_2\right)=\bar{c}_\mathrm{a} k_1 x \left(x+2 k_2\right),
\end{equation}
with $\sigma_{1,2}=\tfrac{1}{2}\left(-\bar{c}_\mathrm{b}\pm\sqrt{\bar{c}_\mathrm{b}^2+4}\right)$. The roots $\sigma_{1,2}$ are both real numbers with $0<\sigma_1\le 1$ and $\sigma_2\le -1$. The case of $\bar{c}_\mathrm{b}=0$ gives $\sigma_{1,2}=\pm 1$. 

Expansion of equation \eqref{eq:factrorized_Pa_2} gives $P=0$ with
\begin{equation}\label{eq:Polynomial_equation}
    P=x^4 + c_3 x^3 + c_2 x^2+ c_1 x+c_0,
\end{equation}
and
\begin{align}
    c_3 &= \bar{c}_\mathrm{b}+k_1,\\
    c_2 &= -\left(1+k_1\left( \bar{c}_\mathrm{a}- \bar{c}_\mathrm{b}-k_2\right)\right),\\
    c_1 &=-k_1\left(1+k_2\left(2\bar{c}_\mathrm{a}-\bar{c}_\mathrm{b}\right)\right),\\
    c_0&=-k_1 k_2,
\end{align}
with $c_4=1$. 

Before finding the roots of the equation $P=0$ it is helpful to analyze the nature of its roots, which is studied by considering the 5-tuple of its coefficients, 
\begin{equation}\label{eq:coefficients_P}
\coef[P]=\left(c_4,c_3,c_2,c_1,c_0\right),
\end{equation}
and its signs
\begin{equation}\label{eq:signs_coefficients_P}
\sgn(\coef[P])=\left(\sgn{c_4},\sgn{c_3},\sgn{c_2},\sgn{c_1},\sgn{c_0}\right).
\end{equation}
It is straightforward to see that $\sgn{c_4}=+$, $\sgn{c_3}=+$, and $\sgn{c_0}=-$. The sign of $c_2$ and $c_1$ requires a careful analysis. There are four possible cases: $\left(\sgn{c_2},\sgn{c_1}\right)$:  $\left(+,+\right)$, $\left(+,-\right)$,
$\left(-,+\right)$, and $\left(-,-\right)$.
The 5-tuple $\sgn(\coef[P])$ can have four possible outcomes: $\left(+,+,+,+,-\right)$, $\left(+,+,+,-,-\right)$, $\left(+,+,-,+,-\right)$, and $\left(+,+,-,-,-\right)$. These 5-tuples display one or three changes of sign along the sequence of elements. Descartes's rule of signs states that the number of positive roots of a polynomial, $P$, is either equal to the number of sign changes of $\sgn(\coef[P])$, or is less than it by an even number. The application of Descartes' rule to $P$ gives either one or three positive roots. It can be proved that the polynomial $P$ has only one positive real root by a careful analysis of equation \eqref{eq:factrorized_Pa_2}. The left hand side of equation \eqref{eq:factrorized_Pa_2} is a fourth degree polynomial $P_L(x)$ with only one positive root $\sigma_1$, one negative root $\sigma_2$ and two roots $\kappa_{1,2}$ that could be either negative or a complex conjugate pair. The right hand side of equation \eqref{eq:factrorized_Pa_2} is an upward parabola $P_R(x)$ with roots at zero and $-2k_2$. The coefficients of the quartic term of $P_L(x)$ and the quadratic term of $P_R$ are both positive numbers, therefore $P_L(x)$ and $P_R(x)$ must tend to infinity as $x$ goes to positive or negative infinity. Since the quartic function grows always faster than the quadratic function, and regarding that $P_L(0)=-k_1 k_1<0$, the polynomials $P_L(x)$ and $P_R(x)$ must be equal at only one positive $x$.

In the Appendix is shown that using Ferrari's method \cite{Dickson1914,Dickson1922}, the quartic equation $P=0$ can be written as an associated resolvent cubic equation $R=0$ with 
\begin{equation}\label{eq:rearranging_of_Pa_5_main}
    R=y^3-c_2 y^2+\left(c_1 c_3-4 c_0\right)y+\left(4c_0c_2-c_0 c_3^2-c_1^2\right).
\end{equation}
The cubic equation $R=0$ can be solved by Cardano's method \cite{Caicedo2023}, for which the change of variable $y=\bar{y}+\frac{c_2}{3}$ is necessary to obtain a depressed cubic equation $R_\mathrm{dc}=0$ with \begin{equation}\label{eq:depressed_cubic_1_main}   R_\mathrm{dc}=\bar{y}^3+\bar{p}\bar{y}+\bar{q},
\end{equation}
where
\begin{align}\label{eq:depressed_cubic_1a_main}
    \bar{p}&=c_1 c_3-\frac{c_2^2}{3}-4c_0,\\
    \bar{q}&=\frac{8 c_0 c_2}{3}+\frac{c_1 c_2 c_3}{3}-\frac{2 c_2^3}{27}-c_1^2-c_0 c_3^2,
\end{align}
and the discriminant of $P$ is $\Delta=-4\bar{p}^3-27\bar{q}^2$ \cite{Dickson1914,Dickson1922}. 

The positive root of $P=0$ is given by three cases depending on the sign of the $\Delta$, and the sign of the functions $\xi_{1,2}$,
\begin{equation}\label{eq:roots_vieta_6_main}
        \xi_{1,2}=-\frac{\bar{q}}{2}\pm \frac{1}{2}\sqrt{-\frac{\Delta}{27}}.
\end{equation}
The quantities $\Delta$, $\xi_1$, and $\xi_2$ are functions of the equilibrium constants, $k_1$ and $k_2$, and the effective concentrations, $\bar{c}_\mathrm{a}$ and $\bar{c}_\mathrm{b}$. Explicitly, the positive root of $P=0$ is given by 
\begin{equation}\label{eq:root_x_main}
x=    \begin{cases}
    x_1,& \Delta>0,\\
    x_1& \Delta<0,\;\xi_1>0,\;\xi_2>0,\\
    x_3,& \Delta<0,\;\xi_1<0,\;\xi_2<0,\\
    x_3,& \Delta<0,\;\xi_1>0,\;\xi_2<0.
    \end{cases}
\end{equation}
The roots $x_1$ and $x_3$ are:
\begin{align}
    x_1&=\tfrac{1}{2}\left(-\left(\tfrac{\bar{c}_\mathrm{b}+k_1}{2}-t_1\right)+\sqrt{\left(\tfrac{\bar{c}_\mathrm{b}+k_1}{2}-t_1\right)^2-2y_1+\tfrac{(\bar{c}_\mathrm{b}+k_1)y_1+2k_1\left(1+(2\bar{c}_\mathrm{a}-\bar{c}_\mathrm{b})k_2\right)}{t_1}}\right),\label{eq:x1}\\
    x_3&=\tfrac{1}{2}\left(-\left(\tfrac{\bar{c}_\mathrm{b}+k_1}{2}+t_1\right)+\sqrt{\left(\tfrac{\bar{c}_\mathrm{b}+k_1}{2}+t_1\right)^2-2y_1-\tfrac{(\bar{c}_\mathrm{b}+k_1)y_1+2k_1\left(1+(2\bar{c}_\mathrm{a}-\bar{c}_\mathrm{b})k_2\right)}{t_1}}\right)\label{eq:x3},
\end{align}
with $y_1$ and $t_1$:
\begin{align}\label{eq:y1_t1}  t_1&=\sqrt{1+\tfrac{1}{4}(\bar{c}_\mathrm{b}+k_1)^2+k_1\left(\bar{c}_\mathrm{a}-\bar{c}_\mathrm{b}-k_2\right)+y_1},\\
y_1&=\bar{y}_1-\tfrac{1+k_1\left(\bar{c}_\mathrm{a}-\bar{c}_\mathrm{b}-k_2\right)}{3},
\end{align}
and $\bar{y}_1$:
\begin{equation}\label{eq:y1_bar_main}
    \bar{y}_1=
    \begin{cases}
    \tfrac{2}{3}\sqrt{1+k_1 Q_1+k_1^2 Q_2}\cos{\left(\tfrac{\theta}{3}\right)},& \Delta>0,\\
    \sqrt[3]{|\xi_1|}+\sqrt[3]{|\xi_2|},& \Delta<0,\;\xi_1>0,\;\xi_2>0,\\
    -(\sqrt[3]{|\xi_1|}+\sqrt[3]{|\xi_2|}), & \Delta<0,\;\xi_1<0,\;\xi_2<0, \\
    \sqrt[3]{|\xi_1|}-\sqrt[3]{|\xi_2|},& \Delta<0,\;\xi_1>0,\;\xi_2<0.
    \end{cases}
\end{equation}

The functions $\theta$, $Q_1$, and $Q_2$ are given in the Appendix. The concentration $x$ for the most common case of diprotic acid, with $k_1>4k_2$, obeys $x=x_1$ for most of the concentrations.  

\subsection{Strong base titration of dissolutions of diprotic acids and its buffer mixtures}\label{subsec:titration_functions}

The titration by a strong base of an acid solution, or an acid buffer solution, can be analyzed by using the effective concentrations $\bar{c}_\mathrm{a}$ and  $\bar{c}_\mathrm{b}$. Recall that the effective concentrations are related to the molar analytical concentrations by $\bar{c}_\mathrm{a}=C_\mathrm{a}/C^{\standardstate}$ and $\bar{c}_\mathrm{b}=C_\mathrm{b}/C^{\standardstate}$. A buffer solution is made of a volume $V_\mathrm{a0}$ of an acid solution with analytical concentration $C_\mathrm{a0}$, and volumes $V_{10}$ and $V_{20}$ of salt solutions with analytical concentrations $C_{10}$ and $C_{20}$, respectively. The total volume of the buffer solution is $V_\mathrm{B0}=V_\mathrm{a0}+V_\mathrm{10}+V_\mathrm{20}$. The case of and acid solution is obtained by using $V_{10}=0$, and $V_{20}=0$, in the volume of the buffer: $V_\mathrm{B0}=V_\mathrm{a0}$. The buffer effective concentrations are given by
\begin{align}    \bar{c}_\mathrm{a0}&=\left(c_\mathrm{a0}V_\mathrm{a0}+s_\mathrm{10}V_\mathrm{10}+s_\mathrm{20}V_\mathrm{20}\right)/V_\mathrm{B0}\label{eq:buffer_ca0},\text{ and}\\
\bar{c}_\mathrm{b0}&=\left(s_\mathrm{10}V_\mathrm{10}+2s_\mathrm{20}V_\mathrm{20}\right)/V_\mathrm{B0}.\label{eq:buffer_cb0}
\end{align}
These expression for the effective concentrations are obtained from the analytical concentrations simply by using the scaling factor $1/C^{\standardstate}$. 

A volume $V_\mathrm{b}$ of a strong base with analytical concentration $C_\mathrm{b0}$ is added to the buffer of volume $V_\mathrm{B0}$ and effective concentrations $\bar{c}_\mathrm{a0}$ and $\bar{c}_\mathrm{b0}$. The addition of the strong base changes the volume of the buffer to $V_\mathrm{B}=V_\mathrm{B0}+V_\mathrm{b}$, and the effective concentrations to the titration effective concentrations
\begin{align}    \bar{c}_\mathrm{a}&=\left(c_\mathrm{a0}V_\mathrm{a0}+s_\mathrm{10}V_\mathrm{10}+s_\mathrm{20}V_\mathrm{20}\right)/V_\mathrm{B}\label{eq:buffer_ca},\text{ and}\\
\bar{c}_\mathrm{b}&=\left(c_\mathrm{b0}V_\mathrm{b}+s_\mathrm{10}V_\mathrm{10}+2s_\mathrm{20}V_\mathrm{20}\right)/V_\mathrm{B}.\label{eq:buffer_cb}
\end{align}
The use of the buffer effective concentrations, equations \eqref{eq:buffer_ca0} and \eqref{eq:buffer_cb0}, in the titration effective concentrations, equations \eqref{eq:buffer_ca} and \eqref{eq:buffer_cb}, gives
\begin{align}
    \frac{\bar{c}_\mathrm{a0}}{\bar{c}_\mathrm{a}}-1&=\frac{V_\mathrm{b}}{V_\mathrm{B0}},\label{eq:buffer_ca1}\text{ and}\\
    \left(\frac{c_\mathrm{b0}}{\bar{c}_\mathrm{b}}-1\right)\frac{V_\mathrm{b}}{V_\mathrm{B0}}&=1-\frac{\bar{c}_\mathrm{b0}}{\bar{c}_\mathrm{b}}.\label{eq:buffer_cb1}
\end{align}
The titration effective concentrations can be combined to obtain an equation for $\bar{c}_\mathrm{b}$ in terms of $\bar{c}_\mathrm{a}$, $c_\mathrm{b0}$ and the buffer effective concentrations,
\begin{equation}\label{eq:buffer_titration_cs}
\bar{c}_\mathrm{b}=\frac{\bar{c}_\mathrm{a}}{\bar{c}_\mathrm{a0}}\left(\bar{c}_\mathrm{b0}-c_\mathrm{b0}\right)+c_\mathrm{b0}.
\end{equation}
This is the equation of a straight line with slope $\left(\bar{c}_\mathrm{b0}-c_\mathrm{b0}\right)/\bar{c}_\mathrm{a0}$ and ordinate intercept $c_\mathrm{b0}$. The slope of this straight line could be be negative, zero, or positive, depending on the concentrations of the buffer, $\bar{c}_\mathrm{b0}=s_\mathrm{10}+2s_\mathrm{20}$, and the titrating base, $c_\mathrm{b0}$. 

In addition to the case of the titration of buffer solutions, this equation can be used for the titration of acid solutions. The case of an acid solution is obtained by taking $\bar{c}_\mathrm{b0}=0$ and $\bar{c}_\mathrm{a0}=c_\mathrm{a0}$, to obtain
\begin{equation}\label{eq:acid_titration_cs}
    \bar{c}_\mathrm{b}=c_\mathrm{b0}-\left(\frac{c_\mathrm{b0}}{c_\mathrm{a0}}\right)\bar{c}_\mathrm{a},   
\end{equation}
where $\bar{c}_\mathrm{a}=c_\mathrm{a}$ and ${\bar{c}_\mathrm{b}=c_\mathrm{b}}$.
This is the equation of a straight line with slope ${-c_\mathrm{b0}/c_\mathrm{a0}}$ and ordinate intercept $c_\mathrm{b0}$. 

Equations \eqref{eq:buffer_titration_cs} and \eqref{eq:acid_titration_cs} describe the titrations as straight line paths on the $(\bar{c}_\mathrm{a},\bar{c}_\mathrm{b})$ plane. The addition of the base, with concentration $c_\mathrm{b0}$, increases $\bar{c}_\mathrm{b}$ along a straight line, from  $\bar{c}_\mathrm{b0}$ to $c_\mathrm{b0}$, meanwhile decreases $\bar{c}_\mathrm{a}$, from $\bar{c}_\mathrm{a0}$ to zero. The use of the contours of constant pH on the $(\bar{c}_\mathrm{a},\bar{c}_\mathrm{b})$ plane and the trajectory described by equation \eqref{eq:buffer_titration_cs}, or \eqref{eq:acid_titration_cs}, give the full description of the titration experiments. However, it is more practical to describe the titration experiment as function of the pH, instead of the added strong base. For this, equation \eqref{eq:charge_balance_z} is used, which relates $x$, the pH, with the effective concentrations $\bar{c}_\mathrm{a}$ and $\bar{c}_\mathrm{b}$. After some rearrangement, equation \eqref{eq:charge_balance_z} gives
\begin{equation}\label{eq:charge_balance_z_2}
    x-\tfrac{1}{x}+\bar{c}_\mathrm{b}=\frac{\bar{c}_\mathrm{a}k_1\left(x + 2 k_2\right)}{x^2+k_1 x + k_1 k_2}.
\end{equation}
The use of the ratio between the effective dimensionless concentrations, $\bar{n}={\bar{c}_\mathrm{b}}/{\bar{c}_\mathrm{a}}$, in equation \eqref{eq:charge_balance_z_2} gives \begin{equation}\label{eq:n_Kalka}
    \bar{n}=\frac{k_1\left(x + 2 k_2\right)}{x^2+k_1 x + k_1 k_2}+\frac{1-x^2}{x \bar{c}_\mathrm{a}}.
\end{equation}
This equation has been reported by Kalka previously \cite{Kalka2021}, and works well in the case of $\bar{c}_\mathrm{a}$ constant. However, in the titration experiment the concentrations $\bar{c}_\mathrm{a}$ and $\bar{c}_\mathrm{b}$ are not constants. The last term of the right hand side of this equation depends on $\bar{c}_\mathrm{a}$. This dependence can be eliminated using equation \eqref{eq:buffer_titration_cs}. For this purpose, equation \eqref{eq:buffer_titration_cs} must be written in terms of $\bar{n}$, 
\begin{equation}\label{eq:n_and_effective_cs}
    \bar{n}=\bar{n}_0+c_\mathrm{b0}\left(\frac{1}{\bar{c}_\mathrm{a}}-\frac{1}{\bar{c}_\mathrm{a0}}\right),
\end{equation}
with $\bar{n}_0={\bar{c}_\mathrm{b0}}/{\bar{c}_\mathrm{a0}}$. Algebraic manipulation of equation \eqref{eq:n_and_effective_cs} gives
\begin{equation}\label{eq:ca_effective}
    \frac{1}{\bar{c}_\mathrm{a}}=\frac{\bar{n}-\bar{n}_0}{c_\mathrm{b0}}+\frac{1}{\bar{c}_\mathrm{a0}},
\end{equation}
which can be used in the last term of equation \eqref{eq:n_Kalka} to obtain an expression for $\bar{n}$ in terms of the pH
\begin{equation}\label{eq:n_buffer}
    \bar{n}=\left(\bar{n}_0-\frac{c_\mathrm{b0}}{\bar{c}_\mathrm{a0}}\right)\frac{P_{\mathrm{a0}}}{P_\mathrm{b0}},
\end{equation}
where $\bar{n}\ge\bar{n}_0$ and
\begin{align}
    P_\mathrm{a0}&=P\left(x=10^{7-\mathrm{pH}},\bar{c}_\mathrm{a}=\frac{\bar{c}_\mathrm{a0} c_\mathrm{b0}}{c_\mathrm{b0}-\bar{c}_\mathrm{b0}},\bar{c}_\mathrm{b}=0\right),\label{eq:Pa0_buffer}\\
    P_\mathrm{b0}&=P\left(x=10^{7-\mathrm{pH}},\bar{c}_\mathrm{a}=0,\bar{c}_\mathrm{b}=c_\mathrm{b0}\right),\label{eq:Pb0_buffer}    
\end{align}
with $P$ the polynomial given by equation \eqref{eq:Polynomial_equation}.
The case of the acid titration is given by considering $\bar{c}_\mathrm{b0}=0$ in equations \eqref{eq:n_buffer}--\eqref{eq:Pb0_buffer}. 

The equivalence points for a diprotic acid occur at $\bar{n}=1,2$ in equation \eqref{eq:n_buffer}. The first equivalence point is given when the acid and base concentrations are equal, $\bar{n}=1$, the second equivalence point happens when the base concentration doubles the acid concentration, $\bar{n}=2$.  Since $\bar{n}(\mathrm{pH})$ must be a monotonically growing function of the pH, it must fulfill the condition $n'(\mathrm{pH})>0$.

\section{Results and discussion }

The validity of equations \eqref{eq:root_x_main}--\eqref{eq:y1_bar_main} has been tested by calculating the pH, at different effective concentrations $\bar{c}_\mathrm{a}$ and $\bar{c}_\mathrm{b}$, of diprotic acids with reported values of p$K_1$ and p$K_2$ for 180 diprotic acids \cite{CRChandbook2007} and comparing them with the numerical solution.
The average absolute error in the pH, $\epsilon_\mathrm{pH}$, at millimolar concentrations $\bar{c}_\mathrm{a}$ and $\bar{c}_\mathrm{b}$, is $\epsilon_\mathrm{pH}\lesssim 10^{-5}$ pH units with a standard deviation  $\lesssim 10^{-4}$ pH units. The small error in the calculated pH is caused mainly by the diprotic acids with $\mathrm{p}K_2-\mathrm{p}K_1<-\log_{10}4$.

\subsection{Aqueous dissolution of a diprotic acid}

\begin{figure}[htb]
    \centering
    \includegraphics[scale=0.75]{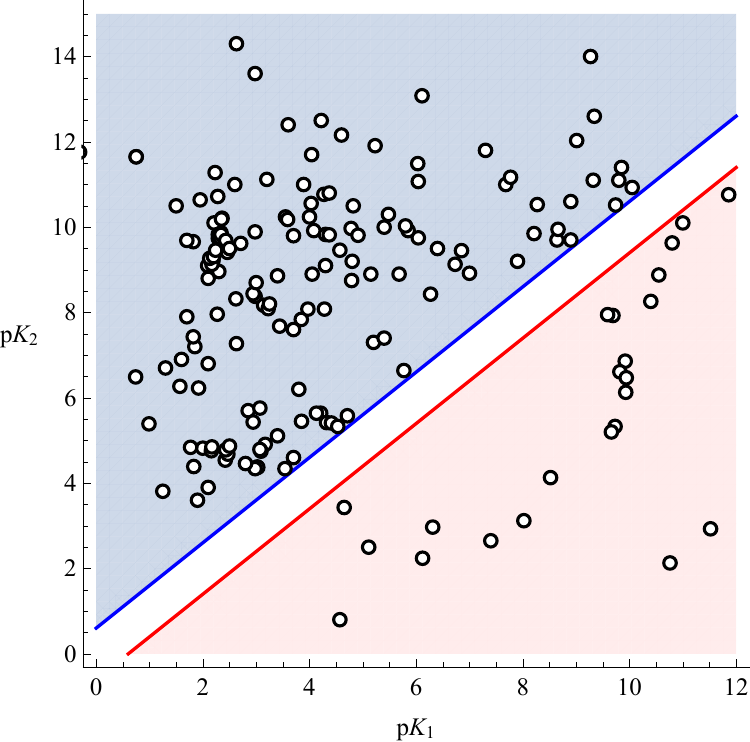}
    \caption{p$K_1$--p$K_2$ plane for a set of diprotic acids in aqueous solution \cite{CRChandbook2007}. The light blue region is given by the condition $\mathrm{p}K_2-\mathrm{p}K_1\ge\log_{10}4$, the light red region is given by the condition $\mathrm{p}K_2-\mathrm{p}K_1\le-\log_{10}4$}
    \label{fig:pK1-pK2}
\end{figure}

The case of a diprotic acid is given by using the conditions $\bar{c}_\mathrm{b}=0$ and $\bar{c}_\mathrm{a}=c_\mathrm{a}$ in equations \eqref{eq:root_x_main}--\eqref{eq:y1_bar_main}. The condition $k_1\ge 4 k_2$, \emph{i.e.},
\begin{equation}\label{eq:pKa_condition}
    \mathrm{p}K_2-\mathrm{p}K_1\ge{\mathrm{log}}_{10}4\approx 0.602,
\end{equation}
makes the discriminant of $P$ a positive number, $\Delta>0$. This is the condition for a quartic equation with four distinct real roots. An inspection of the values of $\mathrm{p}K_1$ and $\mathrm{p}K_2$ of tabulated diprotic weak acids indicates that the condition \eqref{eq:pKa_condition} is fulfilled by many diprotic weak acids \cite{CRChandbook2007,Adams1916,Petrossyan2019}. Figure \ref{fig:pK1-pK2} displays, on the p$K_1$--p$K_2$ plane, the region given by the condition \eqref{eq:pKa_condition} as the light blue region above the blue line. It is clear in the Figure that most of the diprotic weak acids (open black circles) fulfill this condition, however a simple visual inspection of Figure \ref{fig:pK1-pK2} shows that there are weak diprotic acids that fulfill the condition ${\mathrm{p}K_2-\mathrm{p}K_1\le-\log_{10}4}$ (light red region). This condition can be expressed in terms of the acid constants as $k_2/k_1\ge4$, \emph{i.e.} $K_2/K_1\ge4$. Diprotic acids in the light red region have $\mathrm{p}K_1>\mathrm{p}K_2$, examples of these are several nitrogenous organic compounds as Piperazine, Quinine, Phenylbiguanide, $L$-Nicotine, $p$-Benzidine, Sulfamethazine, $m$-Phenylenediamine, $p$-Phenylenediamine, 1,2-Propanediamine, 1,3-Propanediamine, 1,4-Butanediamine, 1,6-Hexanediamine, 1,8-Octanediamine,  \textit{cis}-2,5-Dimethyl\-piperazine, \textit{trans}-1,2-Cyclohexanediamine, \textit{cis}-1,2-Cyclo\-hexane\-diamine, and the alcohol 1,3-Diamino-2-pro\-pa\-nol \cite{CRChandbook2007}.

\subsection{Concentrations $\ce{[H2B]}$, $\ce{[HB^-]}$, and $\ce{[B^{2-}]}$ in terms of $\ce{[H3O+]}$} 

Equation \eqref{eq:factrorized_Pa_2} can be written to give an expression for $\det{\mathsf{K}}$,
\begin{equation}\label{eq:detK_2}
    \det{\mathsf{K}}=\frac{\bar{c}_\mathrm{a} k_1 x \left(x+2 k_2\right)}{\left(x-\sigma_1\right)\left(x-\sigma_2\right)}.
\end{equation}
This equation can be used in equation \eqref{eq:vector_z}, to obtain
\begin{align}
    z_0&=\frac{x \left(x-\sigma_1\right)\left(x-\sigma_2\right)}{k_1(x+2k_2)},\label{eq:z0_k1}\\
    z_1&=\frac{\left(x-\sigma_1\right)\left(x-\sigma_2\right)}{x+2k_2},\label{eq:z1}\\
    z_2&=\frac{k_2\left(x-\sigma_1\right)\left(x-\sigma_2\right)}{x\left(x+2k_2\right)}.\label{eq:z2}
\end{align}
These concentrations are constrained to be positive numbers. Since $0<\sigma_1\le 1$ and $\sigma_2\le -1$, it is necessary to have $x>\sigma_1$. 

It is also possible to have parametric dependence on $\bar{c}_\mathrm{a}$, $\bar{c}_\mathrm{b}$ and $k_2$ by using equation \eqref{eq:c_2}, which gives as result equations \eqref{eq:z1}, \eqref{eq:z2}, and  
\begin{equation}\label{eq:z0_k2}
    z_0=\bar{c}_\mathrm{a}-\frac{\left(x+k_2\right)\left(x-\sigma_1\right)\left(x-\sigma_2\right)}{x\left(x+2k_2\right)}
\end{equation}
instead of \eqref{eq:z0_k1}. The case of a dissolution of the diprotic acid gives $\sigma_1=1$ and $\sigma_2=-1$, with $\bar{c}_\mathrm{b}=0$ and $\bar{c}_\mathrm{a}=c_\mathrm{a}$.

It is convenient to employ logarithmic scaling to describe concentrations and equilibrium constants of highly diluted solutions and weak acids. The p-function of $\mathrm{Q}$ is defined as 
\begin{equation}\label{eq:pQ-function}
\begin{split}
        \mathrm{pQ}&=-\log_{10}a_\mathrm{Q}\\
                   &=-\log_{10}\frac{\gamma_\mathrm{Q}\ce{[Q]}}{C^\circ},
\end{split}
\end{equation}
with $a_\mathrm{Q}$ and $\gamma_\mathrm{Q}$ as the activity and the activity coefficient of $\mathrm{Q}$ respectively \cite{Denbigh1981}. Since equilibrium constants are dimensionless, it is possible to define the p-function of $K$ as $\mathrm{p}K=-\log_{10}K$ \cite{Skoog2022}. 

The case of weak acids, and low concentrations, allows to use the ideal solution approximation, $\gamma_\mathrm{Q}\approx 1$, hence the pH is given by
\begin{equation}
\begin{split}
    \mathrm{pH}&\approx-\log_{10}\frac{\ce{[H3O+]}}{C^\circ}\\
    &\approx-\log_{10}\frac{C^\standardstate x}{C^\circ}\\
    &\approx7-\log_{10}x.
\end{split}
\end{equation}
The p-functions for $\ce{[H_2B]}$, $\ce{[HB^-]}$ and $\ce{[B^{2-}]}$ are given by $\mathrm{pH_2B}\approx7-\log_{10}z_0$, $\mathrm{pHB^{-}}\approx7-\log_{10}z_1$, and $\mathrm{pB^{2-}}\approx 7-\log_{10}z_2$, respectively. The p-function $\mathrm{pH_2B}$ can be expressed in two ways, either using $z_0$ from equation \eqref{eq:z0_k1} to obtain $\mathrm{pH_2B}(k_1,k_2)$, or using $z_0$ from equation \eqref{eq:z0_k2} to get $\mathrm{pH_2B}(c_\mathrm{a},k_2)$. 

\begin{figure}[htb]
    \centering
    \includegraphics{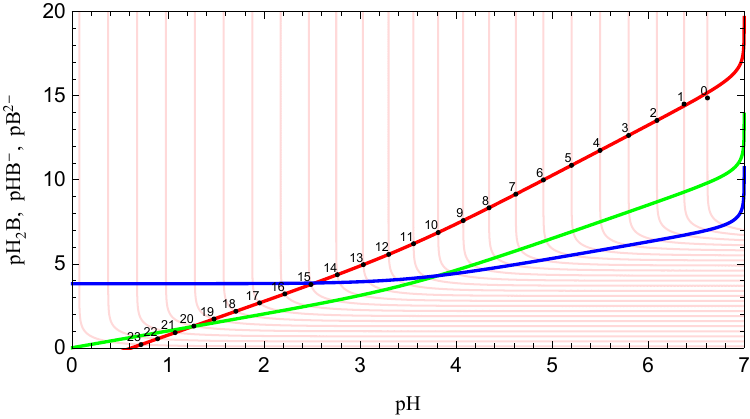}
    \caption{Behaviour of pH$_2$B (red and pink), pHB$^-$ (green) and pB$^{2-}$ (blue) as functions of the pH for oxalic acid, $\ce{H2C2O4}$. Labeled black dots indicate the intersection between pH$_2$B$(k_1,k_2)$ (red) and pH$_2$B$(c_\mathrm{a},k_2)$ (pink) at different concentrations, $c_\mathrm{a}=2^n$, $n=0,1,2, \dots, 23$. Recall that  
    that $C_\mathrm{a}=C^\standardstate c_\mathrm{a}$, hence for $n=0$, $C_\mathrm{a}=10^{-7}\,\mathrm{M}$, and for $n=23$, $C_\mathrm{a}\approx 0.84\,\mathrm{M}$.}
    \label{fig:pZvspH}
\end{figure}

Figure \ref{fig:pZvspH} displays the behaviour of pH$_2$B, pHB$^-$ and pB$^{2-}$ as functions of the pH for different concentrations $c_\mathrm{a}=2^n$ ($n=0,1,\dots,23$) of oxalic acid, $\ce{H2C2O4}$, which has $K_1=5.62\times 10^{-2}$ and $K_2=1.54\times 10^{-4}$ \cite{CRChandbook2007}. In this Figure the intersections between the $\mathrm{pH_2B}(k_1,k_2)$ curve (red) and the $\mathrm{pH_2B}(c_\mathrm{a}, k_2)$ curves (pink) are shown as labeled black points. These intersections give the  pH for the different concentrations $c_\mathrm{a}$. %Notice that $\ce{[H2B]}>\ce{[HB^-]} >\ce{[B^{2-}]}$ only for high concentration, $c_\mathrm{a}>0.1$ and lower pH, $\mathrm{pH}<1.2$. 

\subsection{Use of physical constraints of the system to obtain approximate expressions for $\ce{[H3O+]}$ }

For the diprotic acid, the combined use of equation \eqref{eq:z0_k2} and the condition $z_0>0$, gives the inequality $P_{z_0}<0$ with $P_{z_0}$ given by the monic cubic polynomial
\begin{equation}\label{eq:polynomial_P3}
    P_{z_0}=x^3+\left(k_2-c_\mathrm{a}\right)x^2-\left(1+2c_\mathrm{a}k_2\right)x-k_2.
\end{equation}
Although this polynomial goes to infinity as $x$ goes to infinity, there are values of $x$ for which the inequality $P_{z_0}<0$ is satisfied. This can be seen by analyzing the 4-tuple of $P_{z_0}$ coefficients,
\begin{equation}\label{eq:coeff_Pz0}
\begin{split}
    \coef[P_{z_0}]&=\left(a_3,a_2,a_1,a_0\right)\\
    &=\left(1,k_2-c_\mathrm{a}, -(1+2c_\mathrm{a}k_2),-k_2\right).
\end{split}
\end{equation}
The signs of \eqref{eq:coeff_Pz0} are given by  
\begin{equation}\label{eq:signs_Pz0}
\begin{split}
    \sgn[P_{z_0}]&=\left(\sgn a_3,\sgn a_2,\sgn a_1,\sgn a_0\right)\\
    &=\left(+,\pm,-,-\right),
\end{split}
\end{equation}
Regardless of the value of $\sgn a_2$, there is only one change of sign in \eqref{eq:signs_Pz0}, from positive to negative, for this case Descartes rule  of signs gives that $P_{z_0}$ must have only one positive root. This positive root is a function of $k_2$ and $c_\mathrm{a}$, and gives the upper bound of $x$. Using the method of Caicedo, et al. the upper bound of $x$ is given by
\begin{equation}\label{eq:x_upper}
    x_{\mathrm{ub}}=\tfrac{2}{3}\sqrt{(k_2-c_\mathrm{a})^2+6 c_\mathrm{a} k_2+3}\cos{\left({\theta_{z_0}}/{3}\right)}-\frac{k_2-c_\mathrm{a}}{3},
\end{equation}
with
\begin{align}
    p_{z_0}&=-\tfrac{1}{3}(k_2-c_\mathrm{a})^2-2c_\mathrm{a}k_2-1,\label{eq:p_zeta_0}\\
    q_{z_0}&=\tfrac{2}{27}(k_2-c_\mathrm{a})^3+\tfrac{1}{3}(k_2-c_\mathrm{a})(1+2c_\mathrm{a}k_2)-k_2,\label{eq:q_zeta_0}\\
    \Delta_{z_0}&=-4p_{z_0}^3-27q_{z_0}^2,\label{eq:Delta_zeta_0}\\
    \theta_{z_0}&=\arctan\left(-\frac{q_{z_0}}{2},\frac{\sqrt{\Delta_{z_0}}}{6\sqrt{3}}\right).\label{eq:theta_zeta_0}
\end{align}

The use of Wolfram Mathematica allows to prove that $\Delta_{z_0}>0$ for $c_\mathrm{a}>0$ and $k_2>0$. Since $\Delta_{z_0}>0$, equation \eqref{eq:theta_zeta_0} gives the $0<\theta<\pi$, hence $\cos{(\theta/3)}\ge 1/2$. Furthermore, the use of Wolfram Mathematica allows to find that $\lim_{c_\mathrm{a}\to 0}x_\mathrm{ub}=1$.

It was shown above that the dissociation constants of many diprotic acids are constraint by the condition ${k_1\ge4k_2}$. The use of equations \eqref{eq:ka1} and $\eqref{eq:ka2}$ in the inequality ${k_1\ge4k_2}$ leads to the constraint ${z_1^2\ge4z_0 z_2}$ between the concentrations of the acid and its conjugate bases. The use of equations \eqref{eq:z1}, \eqref{eq:z2} and \eqref{eq:z0_k2} in the inequality ${z_1^2\ge4z_0 z_2}$ gives the inequality $P_z>0$ with
\begin{equation}\label{eq:Polynomial_equation_2}
    P_z=x^3 + 2k_2 x^2 - (1+4c_\mathrm{a}k_2)x-2k_2.
\end{equation}
\begin{figure}[htb]
    \centering
    \includegraphics[scale=0.65]{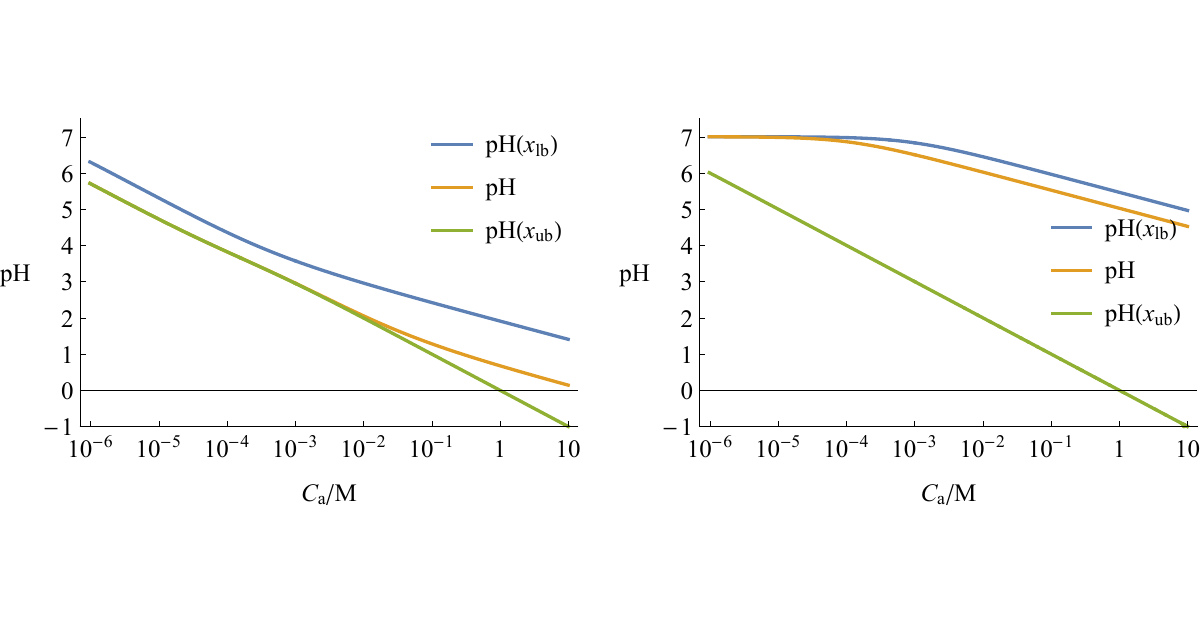}
    \caption{Upper (blue) and lower (green) bounds of the pH for oxalic acid (left) and 1,5-Pentanediamine (right) as functions of the molar concentration $C_\mathrm{a}$. The orange curve represents the exact pH.}
    \label{fig:boundsx}
\end{figure}
This polynomial $P_z$ is the same polynomial for a monoprotic weak acid with $k_\mathrm{a}=2k_2$. Since $\sgn[P_z]=\left(+,+,-,-\right)$, Descartes rule of signs indicates that $P_z=0$ has only one positive root. Using the method of Caicedo et al., this root is given by
\begin{equation}\label{eq:x_lower}
    x_\mathrm{lb}=\tfrac{2}{3}\left(\sqrt{4k_2^2+3c_\mathrm{a}k_2+3}\cos{(\theta_{z}/3)}-k_2\right),
\end{equation}
with
\begin{align}
    p_{z}&=-\tfrac{4k_2^2}{3}-c_\mathrm{a}k_2-1,\label{eq:p_zeta}\\
    q_{z}&=\tfrac{16k_2^3}{27}+\tfrac{2c_\mathrm{a}k_2^2}{3}-\tfrac{k_2}{3},\label{eq:q_zeta}\\
    \Delta_{z}&=-4p_{z}^3-27q_{z}^2,\label{eq:Delta_zeta}\\
    \theta_{z}&=\arctan\left(-\frac{q_{z}}{2},\frac{\sqrt{\Delta_{z}}}{6\sqrt{3}}\right).\label{eq:theta_zeta}
\end{align}
The discriminant $\Delta_z$ is a positive quantity, in fact, by using Wolfram Mathematica it is shown that $\Delta_z\ge 4$. Furthermore using the same software it is shown that $\lim_{c_\mathrm{a}\to 0,k_2\to 0}{x_{\mathrm{lb}}}=1$, and that $\lim_{c_\mathrm{a}\to 0,k_2\to \infty}{x_{\mathrm{lb}}}=1/\sqrt{2}$. 

The lower and upper bounds to the pH are obtained by ${7-\log_{10}{x_{\mathrm{ub}}}}$ and ${7-\log_{10}{x_{\mathrm{lb}}}}$ respectively. Figure \ref{fig:boundsx} displays the lower and upper bounds to the pH as a function of the molar concentration $C_\mathrm{a}$ for oxalic acid (left) and 1,5-Pentanediamine (right). In the case of oxalic acid, the exact pH and the lower pH bound (green) are nearly identical for concentrations $C_\mathrm{a}<10^{-2}\,\mathrm{M}$. On the other hand, for the compound 1,5-Pentanediamine the exact pH and the upper pH bound (blue) are nearly identical for concentrations $C_\mathrm{a}<10^{-4}\,\mathrm{M}$. It is interesting to notice from both panels of Figure \ref{fig:boundsx} that the upper bound pH curve (blue) overstimate the pH for a constant difference for concentrations greater than $10^{-2}\,\mathrm{M}$.

\subsection{Analysis of the dependence of the $\mathrm{pH}$ on p$K_2$}

\begin{figure}[htb]
    \centering
    \includegraphics[scale=0.615]{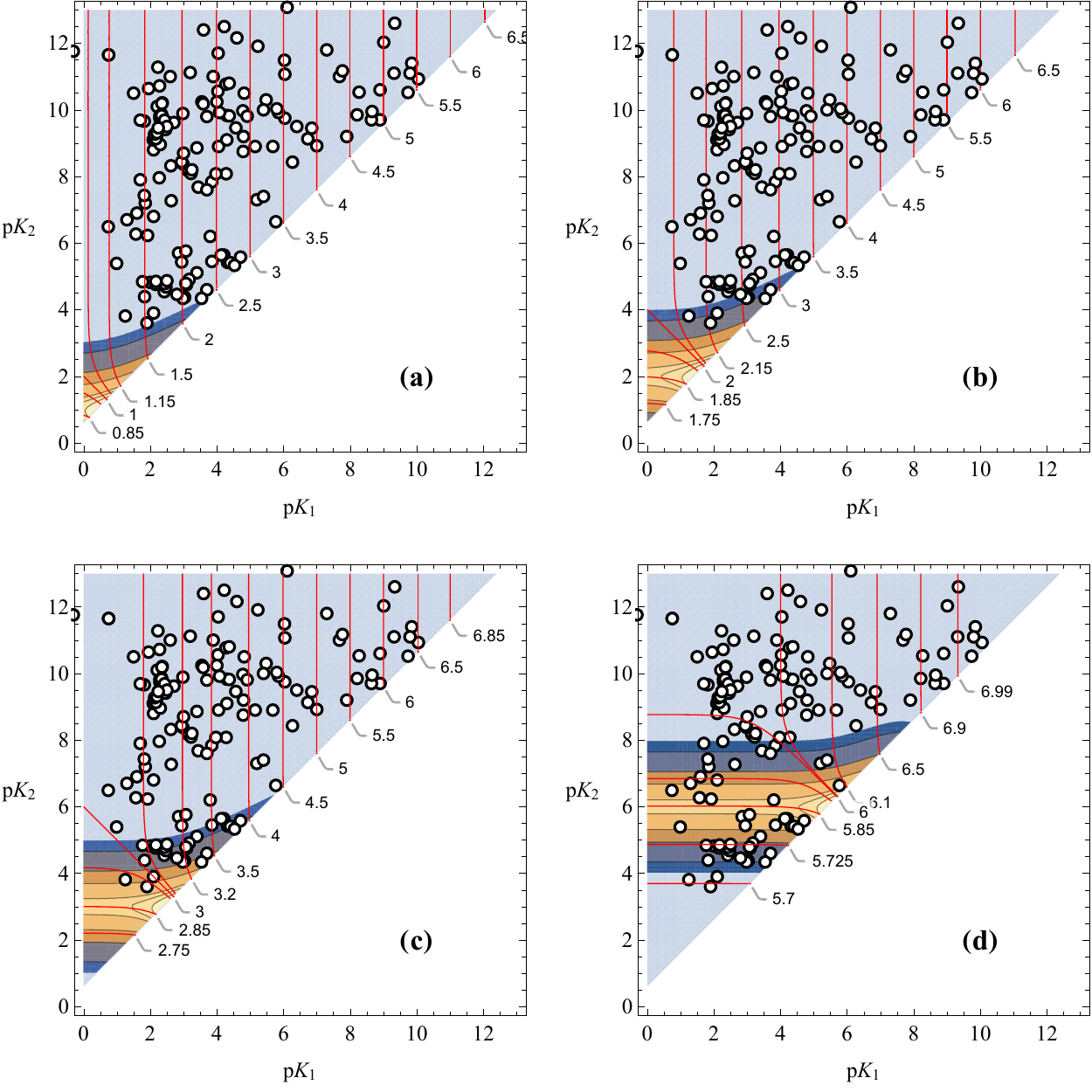}
    \caption{Contours of constant pH (red lines) for different concentrations $C_\mathrm{a}$: (a) $0.1\,\mathrm{M}$; (b) $0.01\,\mathrm{M}$; (c) $10^{-3}\,\mathrm{M}$; (d) $10^{-6}\,\mathrm{M}$. Contours of $\delta\mathrm{pH}_2$ are shown in shading colors from blue to yellow: dark blue 0.01, grey 0.02, dark orange 0.06, orange 0.1, light orange 0.14, dark yellow 0.15, and yellow 0.16. The light blue region has $\delta\mathrm{pH}_2<0.01$. The open circle markers are the values of $\left(\mathrm{p}K_1, \mathrm{p}K_2\right)$ for different weak diprotic acids \cite{CRChandbook2007}.}
    \label{fig:pk1pk2error}
\end{figure}

The pH is calculated by $\mathrm{pH}=7-\log_{10}x$ with $x$ given by equation \eqref{eq:root_x_main}. The partial derivatives
\begin{align}
    \delta\mathrm{pH}_1 &=\left(\frac{\partial\mathrm{pH}}{\partial\mathrm{p}K_1}\right)_{C_\mathrm{a},\mathrm{p}K_2},\label{eq:delta_pH_1}\\
     \delta\mathrm{pH}_2 &=\left(\frac{\partial\mathrm{pH}}{\partial\mathrm{p}K_2}\right)_{C_\mathrm{a},\mathrm{p}K_1},\label{eq:delta_pH_2}
\end{align}
measure how much the pH depends on p$K_1$ or p$K_2$, respectively. 

Figure \ref{fig:pk1pk2error} displays contours of constant  $\delta\mathrm{pH}_2$ as function of p$K_1$ and p$K_2$ for different acid concentrations, $C_\mathrm{a}$:  (a) $0.1\,\mathrm{M}$; (b) $0.01\,\mathrm{M}$; (c) $10^{-3}\,\mathrm{M}$; (d) $10^{-6}\,\mathrm{M}$. Only the acids that fulfill the condition  
$\mathrm{p}K_2\ge\mathrm{p}K_1+\log_{10}4$ are shown. This Figure also shows contours of constant pH as red curves with their respective call-outs indicating the value of the pH. In all the panels, the contours of the derivative $\delta\mathrm{pH}_2$ are shown with contour shading from blue to yellow; the dark blue has $0.01<\delta\mathrm{pH}_2<0.02$, the grey has $0.02<\delta\mathrm{pH}_2<0.06$, the dark orange $0.06<\delta\mathrm{pH}_2<0.1$, the orange $0.1<\delta\mathrm{pH}_2<0.14$, the light orange $0.14<\delta\mathrm{pH}_2<0.15$, the dark yellow $0.14<\delta\mathrm{pH}_2<0.15$, and the yellow $0.16<\delta\mathrm{pH}_2$. 
The open circle markers in all the panels of Figure \ref{fig:pk1pk2error} are the values of $\left(\mathrm{p}K_1, \mathrm{p}K_2\right)$ for different diprotic weak acids \cite{CRChandbook2007}.  
The maximum value of $\delta\mathrm{pH}_2$ is obtained by evaluating $\delta\mathrm{pH}_2$ along the line ${\mathrm{p}K_2=\mathrm{p}K_1+\log_{10}4}$. By doing this a function $\delta\mathrm{p}K_2(\mathrm{p}K_1,C_\mathrm{a})$ is obtained. The use of the function $\mathsf{NMaximize}$ of Wolfram Mathematica gives that ${\max{\left(\delta\mathrm{p}K_2(\mathrm{p}K_1,C_\mathrm{a})\right)}\approx0.17153}$ regardless the values of p$K_1$ and $C_\mathrm{a}$.

Panel (a) of Figure \ref{fig:pk1pk2error} shows that, for $C_\mathrm{a}=0.1\,\mathrm{M}$, p$K_2$ has weak influence on the pH for p$K_1>4$. This is evident by observing that the contours of constant pH are practically vertical lines for $\mathrm{pH}>2.5$ and also by the fact that $\delta\mathrm{pH}_2<0.01$ for the same values of the pH. In the same panel can be observed that the strongest influence of p$K_2$ on the pH, \emph{i.e.}  $0.1<\delta\mathrm{p}K_2\lesssim0.17153$, is seen in the regions with orange and yellow contour shading, and pH$<1.5$. The pH contours in this region are curved instead of straight vertical lines. Panel (a) shows that the approximation of considering the pH independent p$K_2$ is very good for all the acids at a concentration $C_\mathrm{a}=0.1\,\mathrm{M}$. The highest observed value of $\delta\mathrm{pH}_2$ is about 0.17153 units of pH by 1 unit of p$K_2$. This value of $\delta\mathrm{pH}_2$ indicates that a change on 0.5 units of p$K_2$ would produce at most a change of 0.08 units on the pH. This change in the pH is sufficiently small to be within the experimental error, hence, the heuristic approximation of considering the pH dependent only on p$K_1$ is an good approximation at relatively high concentrations of the acid. 

Panels (b) and (c) of Figure \ref{fig:pk1pk2error} show similarities in shape with panel (a). It can be seen that for these concentrations the pH is insensitive to the value of p$K_2$ for p$K_1>5$ and p$K_1>6$, for concentrations $C_\mathrm{a}=10^{-2}$ (b) and $C_\mathrm{a}=10^{-3}$ (c) respectively. Regarding the pH contours, they are insensitive to the p$K_2$ value for pH$>3.5$ and pH$>4.5$ for concentrations $C_\mathrm{a}=10^{-2}$ (b) and $C_\mathrm{a}=10^{-3}$ (c) respectively. Panel (d), $C_\mathrm{a}=10^{-6}\,\mathrm{M}$, shows evident differences with respect to panels (a) to (c). At this low acid concentration the region with $5.73<\mathrm{pH}<6.5$ displays strong sensitivity to the value of p$K_2$, with the strongest effect for pH$\approx5.9$ and p$K_1\approx 6$.   

Panels (a) to (c) of Figure \ref{fig:pk1pk2error} show that the pH contour with $\mathrm{pH}=-\log_{10}C_\mathrm{a}$ is a straight line with negative slope. This line is a boundary between two regions: one with pH contours that are asymptotically independent of p$K_2$ (vertical lines) and other region with pH contours that are asymptotically independent of p$K_1$ (horizontal lines). Although panel (d) does not display this straight line with negative slope, it is clear that there are also regions with asymptotic independence on p$K_1$ and p$K_2$.

\subsection{Strong base titration of diprotic acids and its buffer mixtures}

\begin{figure}[htb]
    \centering
    \includegraphics[scale=0.67]{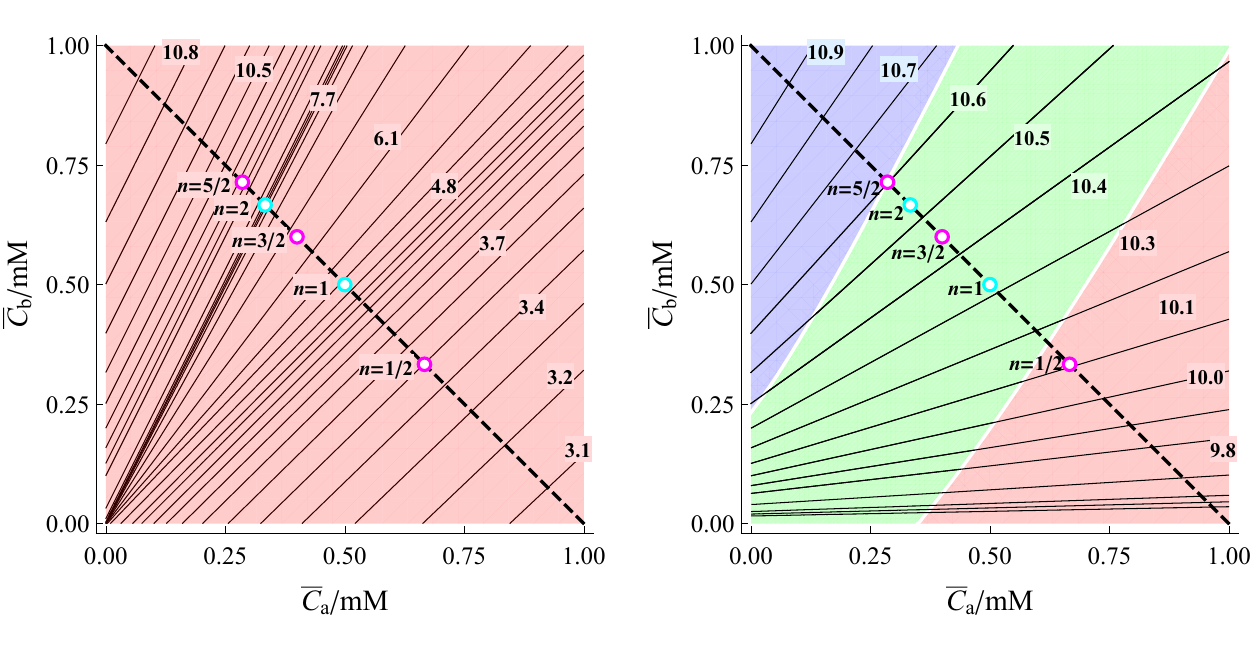}
    \caption{Contours of constant pH on the concentrations plane $(C_\mathrm{a},C_\mathrm{b})$ for diprotic acids. Left: maleic acid, p$K_1=1.92$, p$K_2=6.23$; Right: 1,8-Octanediamine p$K_1=11$, p$K_2=10.1$. The red region is given for $\Delta_\mathrm{dc}>0$, the green region for $\xi_1<0$ and $\xi_2<0$, and the blue $\xi_1>0$ and $\xi_2<0$. The  titration line is shown in dashed on both figures. The equivalence points $n=1,2$ are shown as cyan open circles. The half equivalence points $n=1/2,3/2,5/2$ are shown as magenta open circles.}
    \label{fig:constant_pH_contours}
\end{figure}

Figure \ref{fig:constant_pH_contours} displays the contours of constant pH for maleic acid (left panel) and 1,8-Octane\-diamine (right panel). Maleic acid has p$K_1=1.92$ and p$K_2=6.23$, that is p$K_2-$p$K_1\ge\log_{10}4$, meanwhile the compound 1,8-Octanediamine has p$K_1=11$ and p$K_2=10.1$, and p$K_2-$p$K_1\le -\log_{10}4$. Maleic acid is an example of a pH given uniquely by equation \eqref{eq:x1} with $\bar{y}_1$ given by the first case of equation \eqref{eq:y1_bar_main}, therefore maleic acid displays only one region in the $(C_\mathrm{a},C_\mathrm{b})$ plane. On the other hand the compound 1,8-Octanediamine displays three regions on the $(C_\mathrm{a},C_\mathrm{b})$ plane: the red region is given by using equation \eqref{eq:x1} with $\bar{y}_1$ given by the first case of \eqref{eq:y1_bar_main}, the green region given by equation \eqref{eq:x3} with $\bar{y}_1$ given by the second case of equation \eqref{eq:y1_bar_main}, and the blue region is given by \eqref{eq:x3} with $\bar{y}_1$ given by the third case of equation \eqref{eq:y1_bar_main}. 

\begin{figure}[htb]
    \centering
    \includegraphics[scale=0.67]{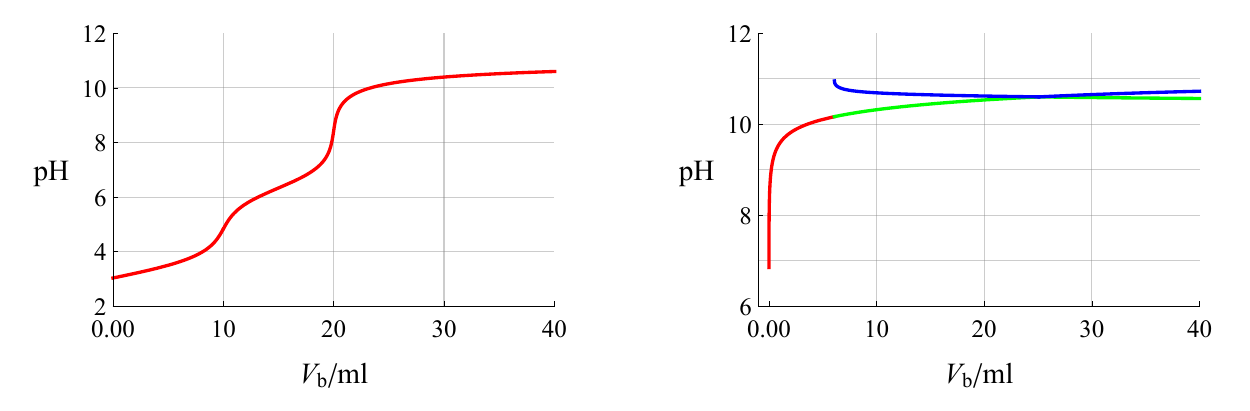}
    \caption{Titration curves for 10$\,\mathrm{ml}$ of diprotic acids at concentration $C_\mathrm{a}=1\,\mathrm{mM}$ using a volume $V_\mathrm{b}$ of strong base with concentration $C_\mathrm{b}=1\,\mathrm{mM}$. (Left) Maleic acid, p$K_1=1.92$, p$K_2=6.23$; (Right) 1,8-Octanediamine p$K_1=11$, p$K_2=10.1$. The red, green and blue curves are calculated using the first, third, and fourth case of equation \eqref{eq:y1_bar_main}, respectively.}
    \label{fig:titration_diprotics}
\end{figure}

Figure \ref{fig:titration_diprotics} shows the titrations curves obtained by adding a volume $V_\mathrm{b}$ of a strong base with concentration $C_\mathrm{b0}=1\,\mathrm{mM}$ to a volume $V_\mathrm{a0}=10\,\mathrm{ml}$ of solution $C_\mathrm{a0}=1\,\mathrm{mM}$ of maleic acid (left) and 1,8-Octanediamine (right). These titration curves are given by the pH along the dashed lines of Figure \ref{fig:constant_pH_contours} for the case $C_\mathrm{a0}=1\,\mathrm{mM}$ and $C_\mathrm{b0}=1\,\mathrm{mM}$. 

The left panel of Figure \ref{fig:titration_diprotics} displays the typical equivalence points for diprotic acids with two equivalence points. The first equivalence point occurs at $V_\mathrm{b}\approx 10\,\mathrm{ml}$ with $\mathrm{pH}\approx 5$, the second equivalence point occurs at $V_\mathrm{b}\approx 20\,\mathrm{ml}$ with $\mathrm{pH}\approx 8$. In contrast, the right panel of Figure \ref{fig:titration_diprotics} does not display equivalence points. The titration curve for 1,8-Octanediamine is made by joining three different titration curves: red, green and blue (from left to right). The initial solution of 1,8-Octanediamine has pH slightly below 7, as the base is added the pH grows rapidly reaching a pH above 10. This behaviour is described by the red curve of Figure \ref{fig:titration_diprotics} (right). As the volume of added base increases, the pH grows from 10 to approximately 10.5, following the green curve. Finally, for $V_\mathrm{b}>25\,\mathrm{ml}$ the titration experiment follows the blue curve reaching a final pH slightly above 10.5 at $V_\mathrm{b}=40\,\mathrm{ml}$.   

\begin{figure}[htb]
    \centering
    \includegraphics[scale=0.67]{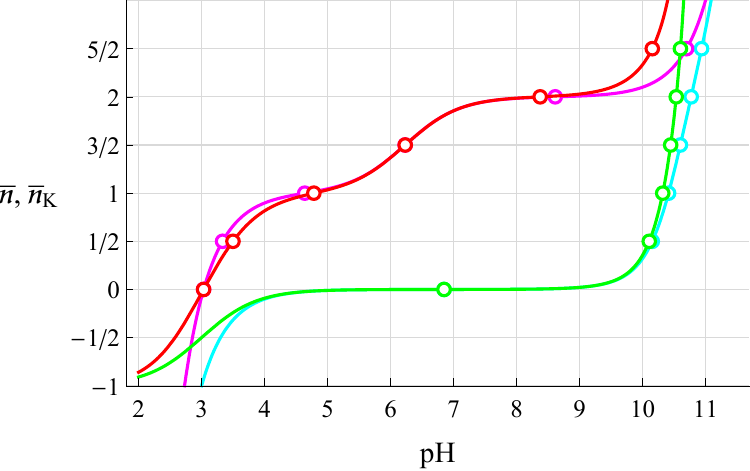}
    \caption{Titration functions $\bar{n}(\mathrm{pH})$, and $\bar{n}_\mathrm{K}$, obtained from equations \eqref{eq:n_buffer} (green and red) and \eqref{eq:n_Kalka} (cyan and magenta), respectively. The first and second equivalence points occur at $\bar{n},\bar{n}_K=1,2$, the half-equivalence points occur at $\bar{n},\bar{n}_\mathrm{K}=1/2, 3/2, 5/2$. Maleic acid (red and magenta) displays the typical titration curve of a diprotic acid, 1,8-Octanediamine (green and cyan) does not diaplay equivalence points.}
    \label{fig:n}
\end{figure}

The use of equation \eqref{eq:n_buffer} allows to obtain the pH at the equivalence points for acid and buffer solutions. The equivalence points $\bar{n}=1, 2$, and the half-equivalence points ${\bar{n}=1/2, 3/2}$, as functions of the  pH, are displayed in Figure \ref{fig:n} for maleic acid, p$K_1=1.92$, p$K_2=6.23$, and for 1,8-Octanediamine, p$K_1=11$, p$K_2=10.1$. It is seen in this Figure that the equivalence point for $\bar{n},\bar{n}_\mathrm{K}=3/2$ are the same for maleic acid, but not for 1,8-Octanediamine. This Figure also shows that the use of equation \eqref{eq:n_Kalka} gives a wrong pH at the equivalence points $\bar{n}=1, 2$. The first equivalence point is shifted to more acid pHs meanwhile the second equivalence point is shifted to basic pHs. 

\subsection{pH stability of buffer solutions} 

\begin{figure}[htb]
    \centering
    \includegraphics[scale=0.67]{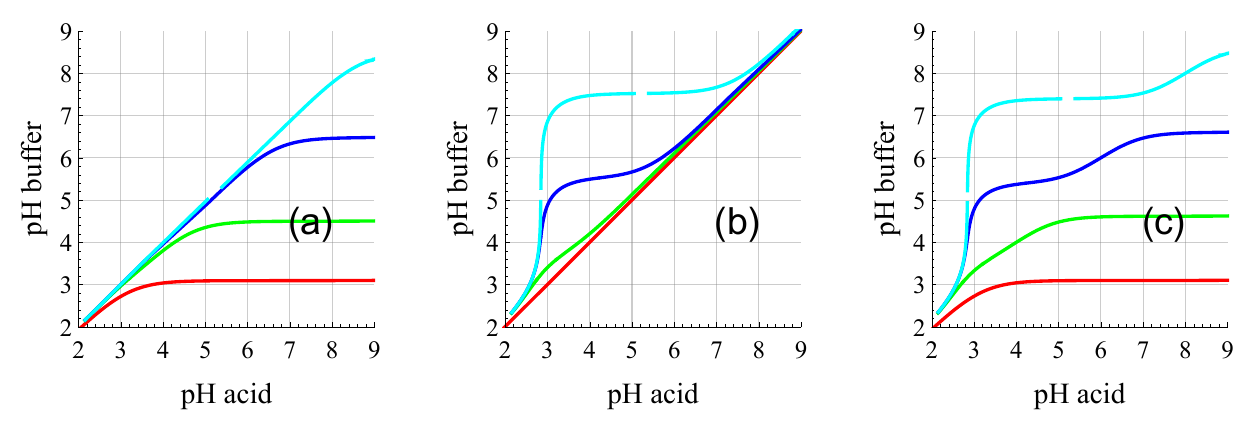}
    \caption{$\beta(V_\mathrm{b})$ curves for buffer solutions of different acids prepared with: (a) only monobasic salt, (b) only dibasic salt, and (c) both monobasic and dibasic salts. All the acids have p$K_1=1$ and: p$K_2=1$ (red), p$K_2=4$ (green), p$K_2=6$ (blue), and p$K_2=8$ (cyan).}
    \label{fig:buffer_stability_row}
\end{figure}

In the titration experiment a volume $V_\mathrm{b}$ of a base solution, with concentration $c_\mathrm{b0}$, is added to a volume $V_{\mathrm{B0}}$ of a buffer solution with concentrations $\bar{c}_{\mathrm{a0}}$ and  $\bar{c}_{\mathrm{b0}}$. The concentrations $\bar{c}_{\mathrm{a}}$ and  $\bar{c}_{\mathrm{b}}$ as functions of $V_\mathrm{b}$ are given by equations \eqref{eq:buffer_ca} and \eqref{eq:buffer_cb}. Although equation \eqref{eq:n_buffer} can be used to analyze the stability of a buffer solution, it is more convenient to use the parametric curve
\begin{equation}\label{eq:buffer_parametric}
\beta(V_\mathrm{b})=\left\{\mathrm{pH}_\mathrm{acid}\left(V_\mathrm{b}\right),\mathrm{pH}_\mathrm{buffer}\left(V_\mathrm{b}\right)\right\},
\end{equation}
where $\mathrm{pH}_\mathrm{acid}\left(V_\mathrm{b}\right)$ is the pH of the acid as function of added base, and $\mathrm{pH}_\mathrm{buffer}\left(V_\mathrm{b}\right)$ is the pH of the buffer as function of added base. Figure \ref{fig:buffer_stability_row} displays $\beta(V_\mathrm{b})$ for acid and buffer solutions prepared with the same number of moles of the acid, $C_\mathrm{a}V_\mathrm{a0}=7.5\times 10^{-3}$ moles, and titrated with the same strong base, $C_\mathrm{b0}=7.5\,\mathrm{mM}$. The buffer solutions of panel (a) are prepared by adding $C_\mathrm{10}V_{10}=2.5\times 10^{-3}$ moles of the monobasic salt only; The buffer solutions of panel (b) are prepared by adding $C_\mathrm{20}V_{20}=2.5\times 10^{-3}$ moles of the dibasic salt only;  The buffer solutions of panel (c) are prepared by adding  $C_\mathrm{10}V_{10}=2.5\times 10^{-3}$ moles of the monobasic salt, and $C_\mathrm{20}V_{20}=2.5\times 10^{-3}$ moles of the dibasic salt. The three panels show four curves for different acids all with p$K_1=1$ and p$K_2=1$ (red), p$K_2=4$ (green), p$K_2=6$ (blue), and p$K_2=8$ (cyan). 

The pH stability of a buffer solution is given when the  $\beta(V_\mathrm{b})$ curve is horizontal, \emph{i.e.} regardless the change in the pH of the acid solution, the pH of the buffer remains stable.  The red curve of panels (a) and (c) of Figure \ref{fig:buffer_stability_row} display the best pH stability. These red curves are produced by acids with p$K_1=1$ and p$K_2=1$. The red  $\beta(V_\mathrm{b})$ curves of panels (a) and (c) display buffer stability at $\mathrm{pH}_\mathrm{buffer}\approx 3$ and $\mathrm{pH}_\mathrm{acid}> 4$. The green  $\beta(V_\mathrm{b})$ curves of panels (a) and (c) display buffer stability at $\mathrm{pH}_\mathrm{buffer}\approx 4.7$ and $\mathrm{pH}_\mathrm{acid}> 6$, for and acid with p$K_1=1$ and p$K_2=4$. The cyan curves of panels (b) and (c) of Figure \ref{fig:buffer_stability_row} show that the dibasic salt produces basic pH stability for acids with higher p$K_2$.  

\section{Declarations}

\subsection{Ethical Approval}
This work does not involve studies in humans and/or animals. There was no need for ethics committee approval. 

\subsection{Competing interests}
The authors declare no competing interests.

\subsection{Authors' contributions}
Juan C. Morales made analytical and numerical calculations. Carlos A. Arango performed analytical and numerical calculations, wrote the manuscript, prepared the figures, and performed the analysis of results.

\subsection{Funding}

This work has been financed by the OMICAS program, Project ID: FP44842-217-2018, and the internal research grants of Universidad Icesi. The OMICAS program acronym stands for ``In-silico Multiscale Optimization of Sustainable Agricultural Crops'', a member of the Scientific Colombia Ecosystem, sponsored by the World Bank, and the Colombian Ministries of Science, Technology and Innovation (Minciencias), Education, Industry and Tourism, and the ICETEX. 

\subsection{Availability of data and materials}

The data and Wolfram Mathematica codes used for this study are available from the corresponding author on request.

\section{Appendix}

\subsection{Mathematical solution of $P=0$}

\subsubsection{The resolvent cubic equation}
The solution of equation $P=0$ by Ferrari's method requires to find its resolvent cubic equation \cite{Dickson1914,Dickson1922}. The standard procedure to obtain the resolvent cubic of a quartic equation begins by writing $P=0$ in its equivalent form 
\begin{equation}\label{eq:rearranging_of_Pa}
\left(x^2+\tfrac{1}{2}c_3x\right)^2=\left(\tfrac{1}{4}c_3^2-c_2\right)x^2-c_1 x - c_0.
\end{equation}
The addition of a quantity $y/2$ inside the squared term of the left hand side, and the addition of the compensation terms on the right hand side gives, after simplification,  
\begin{equation}\label{eq:rearranging_of_Pa_2}
    \left(x^2+\tfrac{1}{2}c_3x+\tfrac{y}{2}\right)^2=\left(\tfrac{1}{4}c_3^2-c_2+y\right)x^2+(\tfrac{1}{2}c_3y-c_1) x - c_0 + \tfrac{1}{4}y^2.
\end{equation}
The left hand side of this equation can be written as a complete square, that is, equation \eqref{eq:rearranging_of_Pa_2} can be written
\begin{equation}\label{eq:rearranging_of_Pa_3}
    \left(t x + \frac{c_3 y - 2 c_1}{4 t}\right)^2=\left(\tfrac{1}{4}c_3^2-c_2+y\right)x^2+(\tfrac{1}{2}c_3y-c_1) x - c_0 + \tfrac{1}{4}y^2,
\end{equation}
with $t=t(y)\ne0$, given that $t^2=\tfrac{1}{4}c_3^2-c_2+y$, and
\begin{equation}\label{eq:rearranging_of_Pa_4}
        \left(\frac{c_3 y - 2c_1}{4t}\right)^2=\tfrac{1}{4}y^2-c_0.
\end{equation}
The expansion of equation \eqref{eq:rearranging_of_Pa_4} gives, after simplification, the resolvent cubic $R=0$, with 
\begin{equation}\label{eq:rearranging_of_Pa_5}
    R=y^3-c_2 y^2+\left(c_1 c_3-4 c_0\right)y+\left(4c_0c_2-c_0 c_3^2-c_1^2\right).
\end{equation}
This equation has three roots $y_i$, $i=1,2,3$. The use of one of these roots in equations \eqref{eq:rearranging_of_Pa_2} and \eqref{eq:rearranging_of_Pa_3} gives 
\begin{equation}\label{eq:rearranging_of_Pa_6}
    \left(x^2+\tfrac{1}{2}c_3 x+\tfrac{y_i}{2}\right)^2=\left(t_i x + \frac{c_3 y_i - 2c_1}{4 t_i}\right)^2,
\end{equation}
with $i=1,2,3$ and $t_i=t(y_i)$. Each of these equations split in two quadratic equations,
\begin{align}
    x^2+\left(\tfrac{1}{2}c_3-t_i\right)x+\tfrac{1}{2}y_i-\frac{c_3 y_i - 2c_1}{4 t_i}&=0,\label{eq:rearranging_of_Pa_7a}\\
    x^2+\left(\tfrac{1}{2}c_3+t_i\right)x+\tfrac{1}{2}y_i+\frac{c_3 y_i - 2c_1}{4 t_i}&=0,\label{eq:rearranging_of_Pa_7b}
\end{align}
with $i=1,2,3$. The roots of $P=0$ satisfy these quadratic equations \cite{Dickson1914,Dickson1922}.
The discriminants of the quartic equation, $\Delta$, and its resolvent cubic equation, $\Delta_\mathrm{rc}$, are identical \cite{Dickson1914,Dickson1922}. The restriction $k_1\ge 4 k_2$ gives that $\Delta>0$, hence $\Delta_\mathrm{rc}>0$ and the resolvent cubic equation \eqref{eq:rearranging_of_Pa_5} must have three distinct real roots. For the case $\Delta_\mathrm{rc}<0$ the cubic $R=0$ has one real root and two non-real complex conjugate roots \cite{Dickson1914,Dickson1922}.

\subsubsection{Solution of the resolvent cubic equation $R=0$}

The third order polynomial equation $R=0$ can be solved by Cardano's method. The change of variable $y=\bar{y}+\frac{c_2}{3}$ gives the depressed cubic equation $R_\mathrm{dc}=0$, with 
\begin{equation}\label{eq:depressed_cubic_1}   R_\mathrm{dc}=\bar{y}^3+\bar{p}\bar{y}+\bar{q},
\end{equation}
and
\begin{align}\label{eq:depressed_cubic_1a}
    \bar{p}&=c_1 c_3-\frac{c_2^2}{3}-4c_0,\\
    \bar{q}&=\frac{8 c_0 c_2}{3}+\frac{c_1 c_2 c_3}{3}-\frac{2 c_2^3}{27}-c_1^2-c_0 c_3^2,
\end{align}
and discriminant $\Delta_\mathrm{dc}=-4\bar{p}^3-27\bar{q}^2$, which is equal to $\Delta$ and $\Delta_{\mathrm{rc}}$ \cite{Dickson1914,Dickson1922}.

The use of Vieta's substitution,
$\bar{y}=\bar{z}-\frac{\bar{p}}{3\bar{z}}$, gives the polynomial equation
\begin{equation}\label{eq:depressed_cubic_2}
    \bar{z}^3-\frac{\bar{p}^3}{27\bar{z}^3}+\bar{q}=0.
\end{equation}
Multiplication of \eqref{eq:depressed_cubic_2} by $\bar{z}^3$ gives
\begin{equation}\label{eq:depressed_cubic_3}
    \bar{z}^6+\bar{q}\bar{z}^3-\frac{\bar{p}^3}{27}=0,
\end{equation}
which is equivalent to the quadratic equation $\xi^2+\bar{q}\xi-\tfrac{\bar{p}^3}{27}=0$, in the variable $\xi=\bar{z}^3$, with roots
\begin{equation}
\begin{split}\label{eq:roots_vieta_6}
        \xi_{1,2}&=-\frac{\bar{q}}{2}\pm\sqrt{\frac{27\bar{q}^2+4\bar{p}^3}{108}}\\
        &=-\frac{\bar{q}}{2}\pm \frac{1}{2}\sqrt{-\frac{\Delta_\mathrm{dc}}{27}}.
\end{split}
\end{equation}
The physical case of diprotic acids with $k_1 \ge 4k_2$ \cite{Adams1916,Petrossyan2019} gives $\Delta_\mathrm{dc}>0$ for $\bar{c}_\mathrm{a}\ge0$ and $\bar{c}_\mathrm{b}\ge0$, therefore $\xi_{1,2}$ are a complex conjugate pair. On the other hand, for the less common case of diprotic acids with $k_1 < 4k_2$ is possible to have $\Delta_{\mathrm{dc}}<0$, hence $\xi_{1,2}$ are real conjugates, on part of the plane $\bar{c}_\mathrm{a}$-$\bar{c}_\mathrm{b}$, with $\xi_1>\xi_2$.

It is convenient to define $\zeta=\xi_1$ and $\zeta^*=\xi_2$ for the case $k_1 \ge 4k_2$ and $\xi=\xi_1$ and $\bar{\xi}=\xi_2$ for the case $k_1 < 4k_2$, with $\zeta^*$ and $\bar{\xi}$ as the complex and real conjugate of $\zeta$ and $\xi$ respectively.

The use the polar representation for $\zeta$ gives
 ${\zeta=\|\zeta\|e^{i\theta}}$ for the case $k_1\ge 4k_2$,  with 
\begin{align}
    \|\zeta\|&=\frac{1}{2}\sqrt{\bar{q}^2+\frac{\Delta_\mathrm{dc}}{27}}=\sqrt{\frac{-\bar{p}^3}{27}},\label{eq:z_modulus_R}\\
    \theta&=\arctan{\left(-\frac{\bar{q}}{2},\frac{\sqrt{\Delta_\mathrm{dc}}}{6\sqrt{3}}\right)}\label{eq:theta_def_R},
\end{align}
with $\theta\in(0,\pi)$ as the angle between $\zeta$ and the positive real axis on the Argand plane. The angle $\theta$ is related to the trigonometric solution obtained by Nickalls for the roots of the cubic equation \cite{Nickalls1993}. The polar representation of $\xi$ and $\bar{\xi}$, case $k_1 < 4k_2$, gives $\xi=|\xi_1|e^{i\theta_1}$ and $\bar{\xi}=|\xi_2|e^{i\theta_2}$ with $\theta_{1,2}=\frac{\pi}{2}(1-\sgn{\xi_{1,2}})$, and $\xi>\bar{\xi}$.

%with few exceptions. Table \ref{tab:Delta_p_k_acids} displays some organic acids with $\mathrm{p}K_2-\mathrm{p}K_1<\log_{10}4$.

%\begin{table}[h]
%    \centering
%    \begin{tabular}{l|c|c}
%       substance & $\mathrm{p}K_1$ & $\mathrm{p}K_2$\\
%       \hline
%       2-aminophenol  & 9.28 & 9.72 \\
%       3-aminophenol  & 9.83 & 9.87 \\
%       D-camphoric acid & 4.57 & 5.10 \\
%       1,3,5-trihydroxybenzene & 8.45 & 8.88 \\
%       Uric acid & 5.40 & 5.53 \\
%       \hline
%    \end{tabular}
%    \caption{$\mathrm{p}K$ values of substances with $\mathrm{p}K_2-\mathrm{p}K_1<\log_{10}4$ at 25\;$^\circ\mathrm{C}$}
%    \label{tab:Delta_p_k_acids}
%\end{table}

%The limit $\lim_{k_2\to 0} \Delta[R_\mathrm{a}]$ is $k_1^2$ times the discriminant of the polynomial for the monoprotic weak acid citep.... 

%\begin{figure}[htb][htb]
%    \centering
%    \includegraphics[scale=0.5]{figure_q.pdf}
%    \caption{Regions of the $k_\mathrm{a}-c_\mathrm{a}$ plane with $q<0$ (gray), and $q>0$ (light gray). The boundary between the regions is given by ${c_\mathrm{a}=\frac{2}{9k_\mathrm{a}}\left(9-k_\mathrm{a}^2\right)}$.}
%    \label{fig:q_positive_negative}
%\end{figure}
The roots $\bar{y}$ of the depressed cubic equation $R_\mathrm{dc}=0$  are given by Cardano's formula
\begin{equation}
    \bar{y}=\alpha+\beta,\label{eq:cardano1}
\end{equation}
where $\alpha=\sqrt[3]{\zeta}$ and $\beta=\sqrt[3]{\zeta^*}$ for $k_1 \ge 4k_2$, and $\alpha=\sqrt[3]{\xi}$ and $\beta=\sqrt[3]{\bar{\xi}}$ for $k_1 < 4k_2$. The cubic roots $\alpha$ and $\beta$ have three values each, $\alpha_n$ and $\beta_n$, with $n=0,1,2$. The combined use of the three roots $\alpha$ and the three roots $\beta$ must give the three roots of $R_\mathrm{dc}=0$. 

The cubic roots $\alpha$ and $\beta$, for the case $k_1 \ge 4k_2$, are given by
\begin{align}
    \alpha_n&=\sqrt[3]{\|\zeta\|}\exp{\left(i\left(\frac{\theta}{3}+\frac{2n\pi}{3}\right)\right)},\\
    \beta_n&=\sqrt[3]{\|\zeta\|}\exp{\left(i\left(-\frac{\theta}{3}+\frac{2n\pi}{3}\right)\right)},
\end{align}
with $n=0,1,2$, and
\begin{equation}
    \sqrt[3]{\|\zeta\|}=\tfrac{1}{3}\sqrt{1+k_1 Q_1+k_1^2 Q_2}.
\end{equation}
for which $Q_1$ and $Q_2$ are
\begin{align}
    Q_1&=-3 k_2 \bar{c}_\mathrm{b}^2+(6 \bar{c}_\mathrm{a}k_2+1)\bar{c}_\mathrm{b}+2\bar{c}_\mathrm{a}-14k_2,\label{eq:Q1}\\
    Q_2&=k_2^2+(4 \bar{c}_\mathrm{a}-\bar{c}_\mathrm{b})k_2+3+(\bar{c}_\mathrm{a}-\bar{c}_\mathrm{b})^2.\label{eq:Q2}
\end{align}

The case $k_1<4k_2$ has 
\begin{align}
    \alpha_n&=\sqrt[3]{|\xi_1|}\exp{\left(i\left(\frac{\theta_1}{3}+\frac{2n\pi}{3}\right)\right)},\\
    \beta_n&=\sqrt[3]{|\xi_2|}\exp\left(i\left(\frac{\theta_2}{3}+\frac{2n\pi}{3}\right)\right),
\end{align}
with $n=0,1,2$.

Since for the case $k_1\ge 4k_2$ the cubic equation $R_\mathrm{dc}=0$ has three real roots, the addition of two cubic roots $\alpha_n+\beta_m$ must give a real number. This is possible only if $\operatorname{Im}(\alpha_n)=-\operatorname{Im}(\beta_m)$.
There are only three possible combinations that fulfill this requirement: 
\begin{align}
    \alpha_0+\beta_0&=2\sqrt[3]{\|\zeta\|}\cos{\left(\tfrac{\theta}{3}\right)},\\
    \alpha_2+\beta_1&=-2\sqrt[3]{\|\zeta\|}\cos{\left(\tfrac{\theta+\pi}{3}\right)},\\
    \alpha_1+\beta_2&=-2\sqrt[3]{\|\zeta\|}\sin{\left(\tfrac{\theta+2\pi}{3}\right)},
\end{align}
which are the roots of $R_{\mathrm{dc}}=0$: $\bar{y}_1$, $\bar{y}_2$, and $\bar{y}_3$, respectively, with $\bar{y}_1>\bar{y}_2>\bar{y}_3$. 

The case $k_1<4k_2$ has only one real solution, and a complex conjugate pair.  Since $\xi>\bar{\xi}$, there are three possibilities: 
\begin{itemize}
\item $\theta_1=\theta_2=0$: the roots are $\alpha_i+\beta_i$ with $i=0,1,2$. The root $\bar{y}_1=\alpha_0+\beta_0$ is the only real solution, $\bar{y}_1=\sqrt[3]{|\xi_1|}+\sqrt[3]{|\xi_2|}$.
\item $\theta_1=\theta_2=\pi$: the roots are $\alpha_i+\beta_i$ with $i=0,1,2$. The root $\bar{y}_1=\alpha_1+\beta_1$ is the only real solution, $\bar{y}_1=-(\sqrt[3]{|\xi_1|}+\sqrt[3]{|\xi_2|})$.
\item $\theta_1=0$, $\theta_2=\pi$: the roots are $\alpha_0+\beta_1$, $\alpha_1+\beta_0$, and $\alpha_2+\beta_2$. The root $\bar{y}_1=\alpha_0+\beta_1$ is the only real solution, $\bar{y}_1=\sqrt[3]{|\xi_1|}-\sqrt[3]{|\xi_2|}$.
\end{itemize}

In summary, the solution $\bar{y}_1$ of the depressed cubic equation is
\begin{equation}\label{eq:y1_bar}
    \bar{y}_1=
    \begin{cases}
    \tfrac{2}{3}\sqrt{1+k_1 Q_1+k_1^2 Q_2}\cos{\left(\tfrac{\theta}{3}\right)},& \Delta_{\mathrm{dc}}>0,\\
    \sqrt[3]{|\xi_1|}+\sqrt[3]{|\xi_2|},& \Delta_{\mathrm{dc}}<0,\;\xi_1>0,\;\xi_2>0,\\
    -(\sqrt[3]{|\xi_1|}+\sqrt[3]{|\xi_2|}), & \Delta_{\mathrm{dc}}<0,\;\xi_1<0,\;\xi_2<0, \\
    \sqrt[3]{|\xi_1|}-\sqrt[3]{|\xi_2|},& \Delta_{\mathrm{dc}}<0,\;\xi_1>0,\;\xi_2<0.
    \end{cases}
\end{equation}

The root $y_1$ of the resultant cubic equation $R=0$ is given by 
\begin{equation}\label{eq:y1}  y_1=\bar{y}_1-\tfrac{1+k_1\left(\bar{c}_\mathrm{a}-\bar{c}_\mathrm{b}-k_2\right)}{3}.\\
\end{equation}

This root is substituted in the quadratic equations \eqref{eq:rearranging_of_Pa_7a} and \eqref{eq:rearranging_of_Pa_7b}, with $t_1=t(y_1)$ given by
\begin{equation}
      t_1=\sqrt{1+\tfrac{1}{4}(\bar{c}_\mathrm{b}+k_1)^2+k_1\left(\bar{c}_\mathrm{a}-\bar{c}_\mathrm{b}-k_2\right)+y_1}.
\end{equation}

The four roots of the quartic equation $P=0$ are given by
\begin{align}\label{eq:roots_of_quartic}
    x_{1,2}&=\tfrac{1}{2}\left(-\left(\tfrac{\bar{c}_\mathrm{b}+k_1}{2}-t_1\right)\pm\sqrt{\left(\tfrac{\bar{c}_\mathrm{b}+k_1}{2}-t_1\right)^2-2y_1+\tfrac{(\bar{c}_\mathrm{b}+k_1)y_1+2k_1\left(1+(2\bar{c}_\mathrm{a}-\bar{c}_\mathrm{b})k_2\right)}{t_1}}\right),\\
    x_{3,4}&=\tfrac{1}{2}\left(-\left(\tfrac{\bar{c}_\mathrm{b}+k_1}{2}+t_1\right)\pm\sqrt{\left(\tfrac{\bar{c}_\mathrm{b}+k_1}{2}+t_1\right)^2-2y_1-\tfrac{(\bar{c}_\mathrm{b}+k_1)y_1+2k_1\left(1+(2\bar{c}_\mathrm{a}-\bar{c}_\mathrm{b})k_2\right)}{t_1}}\right).
\end{align}

Only the roots $x_1$ and $x_3$ can have physical significance, and $x$ is given by
\begin{equation}\label{eq:root_x}
x=    \begin{cases}
    x_1,& \Delta_{\mathrm{dc}}>0,\\
    x_1,& \Delta_{\mathrm{dc}}<0,\;\xi_1>0,\;\xi_2>0,\\
    x_3, & \Delta_{\mathrm{dc}}<0,\;\xi_1<0,\;\xi_2<0, \\
    x_3,& \Delta_{\mathrm{dc}}<0,\;\xi_1>0,\;\xi_2<0.
    \end{cases}
\end{equation}

%%%%%%%%%%%%%%%%%%%%%%%%%%%%%%%%%%%%%%%%%%%%%%%%%%%%%%%%%%%%%%%%%%%%%
%% The "Acknowledgement" section can be given in all manuscript
%% classes.  This should be given within the "acknowledgement"
%% environment, which will make the correct section or running title.
%%%%%%%%%%%%%%%%%%%%%%%%%%%%%%%%%%%%%%%%%%%%%%%%%%%%%%%%%%%%%%%%%%%%%

%%%%%%%%%%%%%%%%%%%%%%%%%%%%%%%%%%%%%%%%%%%%%%%%%%%%%%%%%%%%%%%%%%%%%
%% The same is true for Supporting Information, which should use the
%% suppinfo environment.
%%%%%%%%%%%%%%%%%%%%%%%%%%%%%%%%%%%%%%%%%%%%%%%%%%%%%%%%%%%%%%%%%%%%%

%%%%%%%%%%%%%%%%%%%%%%%%%%%%%%%%%%%%%%%%%%%%%%%%%%%%%%%%%%%%%%%%%%%%%
%% The appropriate \bibliography command should be placed here.
%% Notice that the class file automatically sets \bibliographystyle
%% and also names the section correctly.
%%%%%%%%%%%%%%%%%%%%%%%%%%%%%%%%%%%%%%%%%%%%%%%%%%%%%%%%%%%%%%%%%%%%%
\bibliography{main}

\end{document}